\documentclass{aa}

\usepackage{graphicx}	
\usepackage{ulem}
\usepackage{txfonts}
\usepackage{graphicx}
\usepackage{xcolor}
\usepackage{dsfont}
\usepackage{amsmath}
\usepackage{physics}
\usepackage{siunitx}
\usepackage{tikz}
\usepackage{hyperref} 
\usepackage{acronym}
\usepackage{booktabs}
\begin{document}
\acrodef{CA}{camera assembly}
\acrodef{CalDB}[\texttt{CALDB}]{calibration database}
\acrodef{CalPV}[\texttt{CalPV}]{calibration and phase verification}
\acrodef{CCD}{charge coupled device}
\acrodef{EDR}{eROSITA Early Data Release}
\acrodef{eFEDS}{eROSITA Final Equatorial-Depth Survey}
\acrodef{eRASS1}{eROSITA All-Sky Survey} 
\acrodef{eSASS}[\texttt{eSASS}]{eROSITA Science Analysis Software System}
\acrodef{FFT}{fast Fourier transform}
\acrodef{FOV}{field of view}
\acrodef{GGM}{Gaussian gradient magnitude}
\acrodef{IFT}{information field theory}
\acrodef{ISM}{interstellar medium}
\acrodef{LMC}{Large Magellanic Cloud}
\acrodef{MA}{mirror assembly}
\acrodef{MAP}{maximum a posteriori}
\acrodef{NWR}{noise weighted residual}
\acrodef{PSF}{point spread function}
\acrodef{SN}{supernova}
\acrodef{SNR}{supernova remnant}
\acrodef{S/N}{signal to noise ratio}
\acrodef{TM}{telescope module}
\acrodef{VI}{variational inference}
\acrodef{SRG}{Spectrum-Roentgen-Gamma}
\acrodef{CalPV}{Calibration and Performance Verification}

\title{Bayesian Multiband Imaging of SN1987A in the Large Magellanic Cloud with SRG/eROSITA}
\author{Vincent Eberle\inst{1, 2} \thanks{These authors contributed equally to this work.}, Matteo Guardiani\inst{1, 2} \footnotemark[1], Margret Westerkamp\inst{1, 2} \footnotemark[1], Philipp Frank\inst{1,3}, Michael Freyberg\inst{4}, Mara Salvato\inst{4}, \and Torsten Enßlin\inst{1,2,5}
}
\institute{
  Max Planck Institute for Astrophysics, Karl-Schwarzschild-Straße 1, 85748 Garching, Germany
  \and
  Ludwig-Maximilians-Universität München, Geschwister-Scholl-Platz 1, 80539 Munich, Germany
  \and
  Kavli Institute for Particle Astrophysics \& Cosmology, Stanford University, Stanford, CA 94305, USA
  \and
  Max Planck Institute for Extraterrestrial Physics, Gießenbachstraße 1, 85748 Garching, Germany
  \and 
  Excellence Cluster ORIGINS, Boltzmannstraße 2, 85748 Garching, Germany
}

\date{Received <data>;}

\abstract
{The \ac{EDR} and \ac{eRASS1} data have already revealed a remarkable number of undiscovered X-ray sources. Using Bayesian inference and generative modeling techniques for X-ray imaging, we aim to increase the sensitivity and scientific value of these observations by denoising, deconvolving, and decomposing the X-ray sky. Leveraging information field theory, we can exploit the spatial and spectral correlation structures of the different physical components of the sky with non-parametric priors to enhance the image reconstruction. By incorporating instrumental effects into the forward model, we develop a comprehensive Bayesian imaging algorithm for eROSITA pointing observations. Finally, we apply the developed algorithm to \ac{EDR} data of the \ac{LMC} SN1987A, fusing data sets from observations made by five different telescope modules. The final result is a denoised, deconvolved, and decomposed view of the \ac{LMC}, which enables the analysis of its fine-scale structures, the identification of point sources in this region, and enhanced calibration for future work.}

% context handling (optional)





% maximum of 6 keywords

\titlerunning{Bayesian Multiband Imaging of SN1987A in the LMC with SRG/eROSITA}

\authorrunning{V. Eberle \& M. Guardiani \& M. Westerkamp et. al}

\maketitle
\section{Introduction}
The eROSITA X-ray Telescope~\citep{Predehl2020} on \ac{SRG}~\citep{Sunyaev2021} was launched on July 13th, 2019 from the Baikonour Cosmodrome. Since the full-sky survey of ROSAT  \citep{Truemper:1982} in 1990, eROSITA is the first X-ray observatory to perform a full-sky survey with higher resolution and a larger effective area.
After a \ac{CalPV} phase of pointed and field-scan observations, the main phase of the mission is devoted to multiple all-sky surveys of the celestial sphere, each lasting about 6 months. The amount of data collected by the X-ray observatory in its about 4.3 completed all-sky surveys already has a huge scientific impact.
In order to make use of scientific data, nuisance effects of the instrument need to be understood and removed whenever possible. Amongst others, Poisson noise and the \ac{PSF} of the optical system cause problems to source detection algorithms. Unfortunately, some of these effects are not analytically invertible and thus leave us with an ill-posed problem at hand. 
In this work, we make use of \Acf{IFT} \citep{ensslin09} as a theoretical framework to tackle these problems. The use of prior knowledge and generative modeling enables us to remove instrumental effects, decompose the sky into astrophysical emission components, potentially remove the high-energy particle background, and leave us with an approximation of the posterior distribution.
This permits us to gain knowledge about any posterior measure of interest, such as the mean and the uncertainty of the measured physical quantities.

\subsection{Related work}
X-ray astronomy has developed rapidly since its beginnings in the 1960s, driven by major X-ray missions such as Einstein and ROSAT. This rapid progress has been fueled not only by advancements in instrumentation, with ever-improving telescopes such as Chandra \citep{Chandra}, XMM-Newton \citep{XMM}, and more recently eROSITA, but also by simultaneous developments in imaging techniques. These advancements have steadily increased the amount of information extracted from observations and enabled researchers to address various data-analysis challenges. For instance, tasks such as source detection and the coherent fusion of overlapping datasets -- some of the most difficult tasks in astrophysical imaging -- along with the denoising and deconvolution of X-ray data affected by Poisson noise, have become more manageable due to these innovations.

In \cite{westerkamp24}, an overview of general source detection algorithms is given, such as the sliding cell algorithm algorithm \citep{2001ASPC..238..443C}, the wavelet detection algorithm \citep{Freeman_2002} and the Voronoi tessellation and percolation algorithm \citep{PhysRevE.47.704}, as well as an overview on data fusion techniques currently used and implemented for Chandra data. 
A summary of the data processing and imaging pipelines for the Chandra and XMM-Newton X-ray telescopes is available at \cite{Seward_Charles_2010}. 
This section provides an overview of the state of the art in X-ray imaging techniques specifically for eROSITA. 
For eROSITA data analysis, there is the \ac{eSASS} \citep{Brunner2018,merloni_srgerosita_2024}, which includes all the functionalities of the standard eROSITA processing pipeline, such as event processing, event file and image generation, background estimation and point source detection, and source-specific output such as light curve and spectrum generation. 
In \cite{Brunner_2022}, the standard eROSITA source detection pipeline using \ac{eSASS} for \ac{eFEDS} is elaborated step by step. First, the standard source detection requires a preliminary source list containing all possible source candidates, which is generated using the sliding cell algorithm. 
Based on the preliminary source catalog, the X-ray data is compared to an \ac{PSF} model using maximum likelihood fitting. Finally, as noted in \cite{merloni_srgerosita_2024}, circular regions of appropriate radius can be placed around point sources to exclude them, thereby obtaining point-source-subtracted images.
All the necessary functionalities, including the sliding cell algorithm, are implemented in the corresponding \ac{eSASS} package. 
To test the completeness and accuracy of this source detection pipeline, \cite{Liu_2022} simulated \ac{eFEDS} data, given a specific catalog and background, and applied the source detection algorithm to it. 
\citet{seppi} also estimated the fraction of spurious sources for eRASS1 using the eSASS pipeline in different configurations.

Recently, \cite{merloni_srgerosita_2024} published a catalog of point sources and extended sources in the western Galactic hemisphere using the first of the all-sky scans of \ac{eRASS1}.
In this study, we focus on the imaging of the LMC using EDR data from the \ac{CalPV} phase.
As the nearest star-forming galaxy, the LMC has already been observed and analyzed in its various parts across the entire electromagnetic spectrum, as noted in \cite{zangrandi_first_2024, zanardo_high-resolution_2013}. 
Among other things, the numerous \acp{SNR} present in it are of interest, as studied in \cite{zangrandi_first_2024} on data from eRASS:4, including all data from the eROSITA all-sky surveys eRASS1-4.
To enhance the edges of the shocked gas in the \acp{SNR}, they used the \ac{GGM} filter \citep{sanders_detecting_2016}, resulting in 78 \acp{SNR} and 45 candidates in the \ac{LMC}.
The most famous \ac{SN} in the \ac{LMC} is SN1987A, as the only nearby core-collapse \ac{SN}. SN1987A provides a perfect opportunity to study the evolution of young Type II SNe into the \ac{SNR} stage. 
It has therefore been the subject of several publications and observed by several instruments, including ATCA \citep{zanardo_high-resolution_2013}, XMM-Newton \citep{haberl_xmm-newton_2006}, Chandra \citep{burrows_x-ray_2000}, and recently JWST \citep{Matsuura:2024}.

In this study, we focus on Bayesian imaging methods for X-ray astronomy based on the algorithm D3PO \citep{Selig_2015}, which implements denoising, deconvolution, and decomposition of count data. Decomposition means that, in addition to the total photon flux, the composition of the flux at each pixel is reconstructed using assumptions about prior statistics. The algorithm has been extended by \cite{Pumpe_2018} to reconstruct and decompose multi-domain knowledge. The developed algorithm has been applied to Fermi data to reconstruct the spatio-spectral gamma-ray sky in \cite{Scheel_Platz_2023}, and its capabilities have been shown on X-ray photon-count data for Chandra in \cite{westerkamp24}. Here, we build a novel likelihood model for the eROSITA instrument and advance the prior model for the X-ray sky to reconstruct \ac{LMC} features from \ac{EDR} eROSITA data, as shown below. Moreover, we use \ac{VI} to approximate the posterior instead of the \ac{MAP} approach followed in \cite{Pumpe_2018}.

\section{Observations}\label{sec:data}
We have employed observations from the \ac{EDR} of the \ac{LMC} SN1987A containing  data from the \ac{CalPV} phase of eROSITA. We have used data of the LMC SN1987A from eROSITA in pointing mode with the observation ID 700161 \footnote{The data used are publicly available at \url{https://erosita.mpe.mpg.de/edr/eROSITAObservations/}.}. In total, this observation of the \ac{LMC} includes all seven \acp{TM} of eROSITA. However, we have chosen only to use data from five of these \acp{TM}, specifically TM1, TM2, TM3, TM4 and TM6 (together usually referred to as TM8), since TM5 and TM7, as noted in \cite{merloni_srgerosita_2024}, do not have an on-chip optical blocking filter and suffer from an optical light leak \citep{Predehl2020}. 
The raw data were processed using the \ac{eSASS} pipeline \citep{Brunner_2022} and binned into $1024 \times 1024$ spatial bins and 3 energy bins, 0.2-1.0 keV (red), 1.0-2.0 keV (green) and 2.0-4.5 keV (blue), according to the binning used by Haberl et al.~\footnote{The first light \ac{EDR} image of LMC SN1987A by F. Haberl et al. is shown in \url{https://www.dlr.de/de/aktuelles/nachrichten/2019/04/20191022_first-light-erosita}.}.

\begin{figure}[!h]
  \centering
  \includegraphics[width=\linewidth]{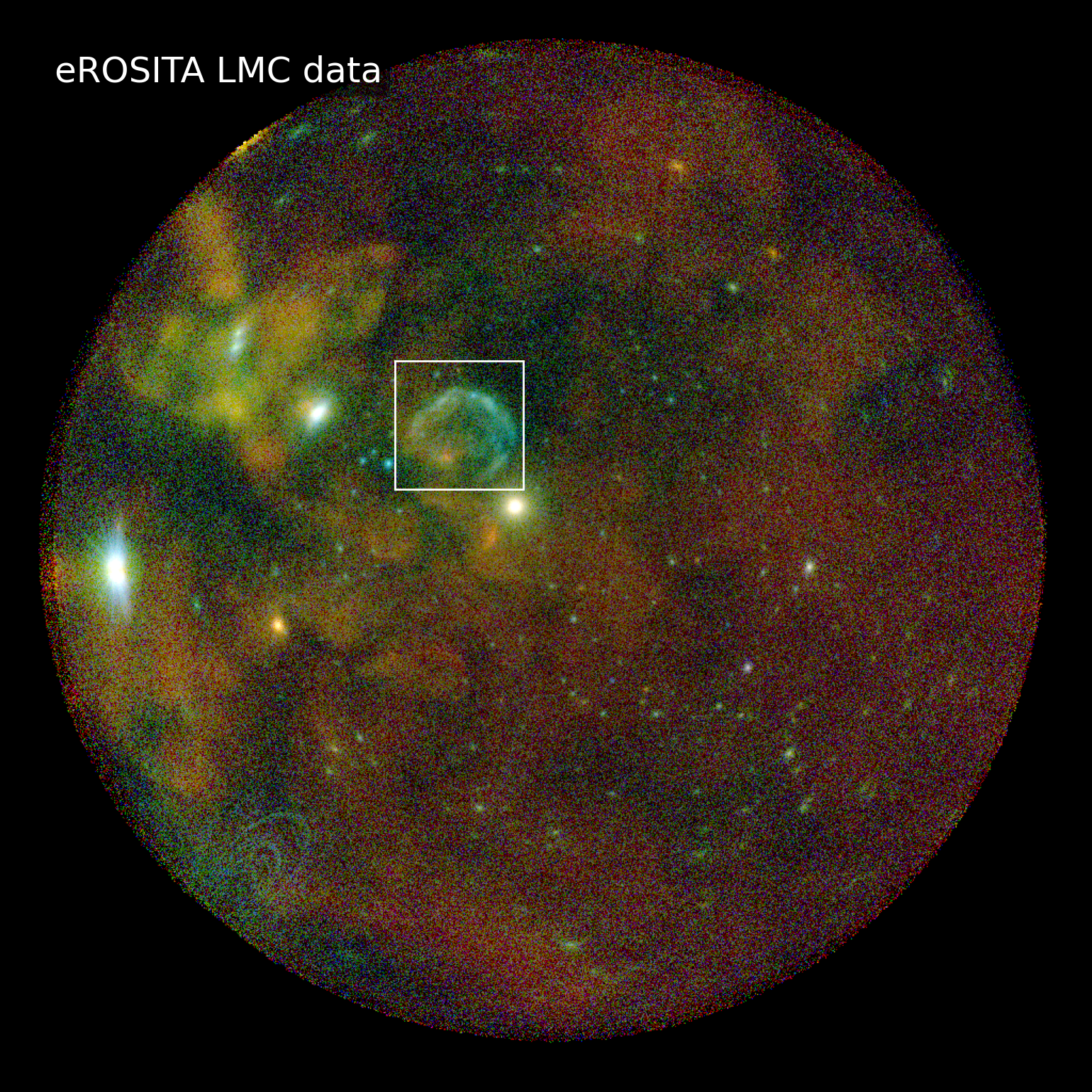}
  \caption{Visualization of the exposure corrected eROSITA EDR TM8 data of ObsID 700161 for three energy bins. Red: 0.2-1.0 keV, Green: 1.0-2.0 keV, Blue: 2.0-4.5 keV. The white box marks the region of 30 Doradus C further discussed in Sect.~\ref{subsec:prior_models}. The bright central point-like source underneath the bottom-right corner of the 30 Doradus C box is SN1987A.}
   \label{fig:Data}
\end{figure}

In particular, we used \texttt{evtool} to generate the cleaned event files and \texttt{expmap} to generate the corresponding exposure maps for each \ac{TM}. 
The specific configurations for \texttt{evtool} and \texttt{expmap} can be found in the Appendix \ref{app:data}. We created new detector maps that included the bad pixels in order to exclude those from the inference. Figure \ref{fig:Data} shows the corresponding RGB image of the eROSITA \ac{LMC} data, where one image pixel corresponds to four arc seconds.  In the appendix in Fig. \ref{fig:TMeROSITAdata} the data per energy bin and \ac{TM} are shown. Figure \ref{fig:exposures} shows the exposures summed over the \acp{TM}.

\section{Methods}\label{sec:methods}
In this section, we present the methods used for X-ray imaging with eROSITA. In the end we want to reconstruct a signal $s$, in our case the X-ray photon flux density field in units of $[1/(\arcsec^2 \times \  \mathrm{s})]$\footnote{To convert the reported fluxes to units of $[\mathrm{keV}/(\mathrm{arcsec}^2 \times \mathrm{s})]$, multiply by $\langle \mathcal{E} \rangle_i$, the average photon energy in keV for each sky reconstruction bin $i$. The resulting value corresponds then to the mean integrated flux in this energy bin.}. The signal is described by a physical field and is a function of spatial coordinates, $x \in \mathbb{R}^2$, and a spectral coordinates, $y= \log(\mathcal{E}/\mathcal{E}_0) \in \mathbb{R}$, where $\mathcal{E}$ is the energy and $\mathcal{E}_0$ the reference energy. In the following, we describe the Bayesian inference of the signal field and its components, in other words the prior and the likelihood model.

\subsection{Imaging with information field theory}\label{subsec:ift_imaging}
X-ray imaging poses a series of different challenges. Astrophysical sources emit photons at a certain rate. 
This rate can be mathematically modeled by a scalar field that varies across the \ac{FOV}, energy, and time. 
After being bent through the instrument's optics, this radiation is then collected by the \acp{CCD}, which records individual photon counts as events. 
This way, the physical information contained in the sources' flux spatio-spectro-temporal distribution is degraded into the observational data.
The mathematical object of a field with an infinite number of degrees of freedom, which is well suited to describe the original flux-rate signal, is not suited to describe a finite collection of event counts.
Recovering the infinite degrees of freedom of the signal field from finite data is a challenging problem that requires additional information. 
\ac{IFT} \citep{ensslin09,ensslin19} provides a mathematical framework to introduce these additional components and solve the inverse problem of recovering fields from data.
The additional information introduced characterizes typical source types found in astrophysical observations, such as point sources, which can be bright but are spatially sparse; diffuse emission, which is nearly ubiquitous across the \ac{FOV} and spatially correlated; and extended sources, which are finite regions of diffuse emission with their own specific correlation structures.
In the context of X-ray imaging, this allows to accurately and robustly reconstruct the underlying photon flux field as the sum of all modeled emission fields.
In essence, upon denoting the quantity of interest, in our case the X-ray flux, with $s$ for signal, we can use prior information on the distribution of $s$, $\mathcal{P}(s)$, to obtain posterior information $\mathcal{P}(s|d)$ on the signal constrained to the observed data $d$ using Bayes' theorem
\begin{align}\label{eq:bayes}
	\mathcal{P}(s|d) = \frac{\mathcal{P}(d|s)\, \mathcal{P}(s)}{\mathcal{P}(d)}.
\end{align}
Here, $\mathcal{P}(d|s)$ is called the likelihood and incorporates information about the instrument's response and noise statistics, while $\mathcal{P}(d)$ is called the evidence and ensures proper normalization of the posterior $\mathcal{P}(s|d)$.
In the following, we will discuss our choices for the prior distribution (Sect.~\ref{subsec:prior_models}), describe how to build the likelihood model which takes into account eROSITA-specific instrumental effects (Sect.~\ref{subsec:likelihood}), and explain how to combine our likelihood and prior models to numerically approximate the posterior distribution as this turns out to be analytically intractable (Sect.~\ref{subsec:inference}). The corresponding models are built using the software package \texttt{J-UBIK} \citep{JUBIK:2024}, the JAX-accelerated universal Bayesian imaging kit, which is based on \texttt{NIFTy.re} \citep{Edenhofer:2024} as a JAX-accelerated version of \texttt{NIFTy} \citep{nifty_selig, Arras:2019}.
Although we focus on eROSITA imaging in this work, the presented algorithm is general and applicable to other photon-count observatories. For instance, in \citet{westerkamp24}, a similar technique is applied to Chandra data. As instrument models are made publicly available through \texttt{J-UBIK}, with plans to expand the included instruments in the future, this framework enables accurate, high-resolution imaging and has the potential to support multi-messenger imaging.

\subsection{Prior models}\label{subsec:prior_models}
Prior models are an essential part of Bayesian inference, allowing us to  infer a field with a virtually infinite number of degrees of freedom from a finite number of data points. Here we explain how we mathematically model different sky components, their underlying assumptions and justifications, and how these models are implemented in a generative way. 

Our signal $s$ is composed of a set of sky components $\{s_i\}$,
\begin{align}
    s = \sum_i s_i \ ,
\end{align}
that differ in their morphology. In this study, these are in particular the point source emission, $s_p$, and the diffuse extended source emission, $s_d$. Building individual prior models for each of these components allows us to decompose the reconstructed, denoised and deconvolved sky into its various sources. The prior models for each of the sky components are implemented as generative models as introduced in \cite{knollmueller2020metric} using the reparametrization trick of \cite{kingma2015variational}. In other words, each of the prior models is described by a set of normal or log-normal models, leading to the final generative model defined via Gaussian processes via inverse transform sampling. In this study, we distinguish between spatially correlated sources, which describe diffuse emission, and spatially uncorrelated sources, which model point sources. For each of the components we have a correlated spectral direction.

There are several ways to implement the correlation in the spatial or spectral dimension. To model the two-dimensional spatial correlation in diffuse emission, we use the correlated field model introduced in \cite{Arras_2022}. In this particular case the two-dimensional field, which we call $\varphi_{\text{ln}} =e^\tau$, is modelled by a log-normal process with $\tau$ being normal distributed, $\mathcal{P}(\tau|T) = \mathcal{N}(\tau, T)$, with unknown covariance $T$,
\begin{align}
    \varphi_{\text{ln}} = e^\tau = e^{A \xi_\tau} ~~ \text{with} ~~ 
    T = AA^\dagger, ~\xi_\tau \sim \mathcal{N}(\xi_\tau, \mathds{1}) \ ,
    \label{eq:gauss_process}
\end{align}
where we denote the Gaussian distribution for a random variable $x$ with covariance $X$ as
\begin{align}
    \mathcal{N}(x,X) := \frac{1}{|2\pi X|^\frac{1}{2}} \exp \biggl( -\frac{1}{2} x^\dagger X^{-1} x \biggr).
\end{align}
Assuming a-priori statistical homogeneity and isotropy, the correlation structure encoded in $T$ can be represented by its power spectrum according to the Wiener-Khinchin theorem. In order to learn the power spectrum and thus the correlation structure simultaneously with the diffuse sky realization, it is implemented by an integrated Wiener process whose parameters are themselves represented by log-normal and Gaussian processes and can thus be learned from the data. For more details on the correlated field model see \cite{Arras_2022}.

For the point sources, we want the two-dimensional spatial field, $\varphi_{\text{ig}}$,  to be pixel-wise uncorrelated, or in other words we want each pixel to be independent. Statistically this is described by a probability distribution, which factorizes in spatial direction. Moreover, we aim for a few bright point sources. As shown in \cite{Guglielmetti_2009} an appropriate probability distribution is the inverse gamma distribution, i.e.
\begin{align}
    \mathcal{P}(\varphi_{\text{ig}}) = \prod_x \Gamma^{-1}\qty(\varphi_{\text{ig}}(x)) \ .
\end{align}
As we aim to perform a spatio-spectral reconstruction of the eROSITA X-ray sky, we add a spectral axis. In this study we consider a power-law behaviour, described by the spectral index $\alpha$ in spectral direction. For the diffuse emission we assume that the spectral index, $\alpha_d$, is spatially correlated, while it is assumed to be spatially uncorrelated for the point source emission $\alpha_p$. This leads to the mathematical definition of the individual components, $s_p$ and $s_d$;
\begin{align}
    s_p(x,y) = \varphi_{\text{ig}}(x) \times e^{\alpha_p(x)\,y } ~~ \text{and} ~~  
    s_d(x,y) = \varphi_{\text{ln}}(x) \times e^{\alpha_d(x)\,y }
\end{align}
In Fig. \ref{fig:Data} it can be seen that both the correlation structure and the spectral power law behaviour in the region of 30 Doradus C are fundamentally different from the diffuse structures that are otherwise present in the data. For the diffuse structures in the \ac{LMC} we expect long correlation structures and a steep power-law slope in the energy direction. 30 Doradus C, on the other hand, has a flat power-law and a shorter correlation length. To account for this, we add another prior component, $s_b$, in the region of a box, $b$, around 30 Doradus C, which has a correspondingly flatter power law and allows for smaller structures, giving us a third component,
\begin{align}
s_b(x,y) = 
\begin{cases} 
\varphi_{\text{ln, b}}(x) \times e^{\alpha_b(x)\,y } & \text{if } x \in b \\
0 & \text{otherwise}
\end{cases} \ .
\end{align}
\begin{figure}[!h]
  \centering
  \includegraphics[width=\linewidth]{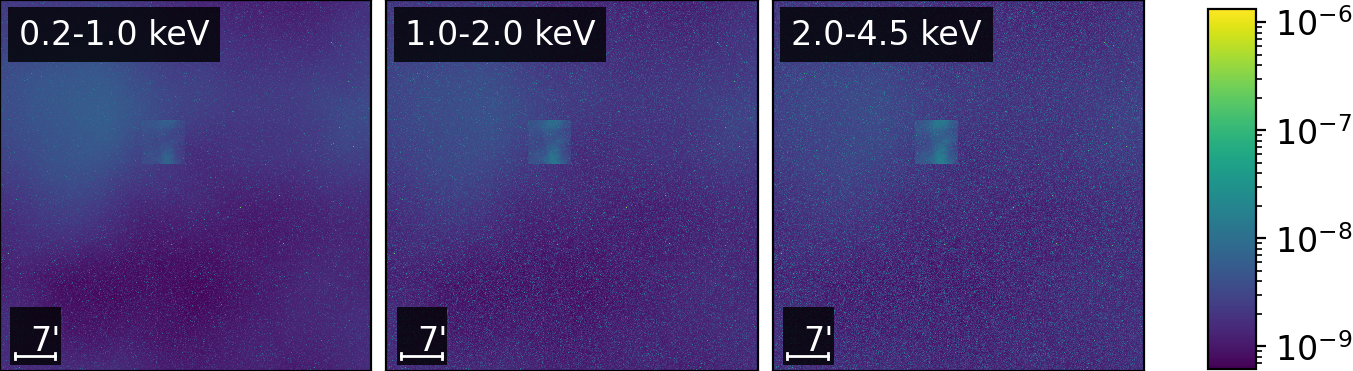}
  \caption{Visualization of one prior sample drawn from the prior model described in Sect.~\ref{subsec:prior_models} for three energy bins in $[1/(\arcsec^2 \times \  \mathrm{s})]$.}
   \label{fig:Priorsample}
\end{figure}
The prior model does not take temporal variation into account and thus assumes time invariant flux.
Fig.~\ref{fig:Priorsample} shows one prior sample drawn from the here described prior model for three energy bins. 
In the Appendix~\ref{app:hyperparams}, we explain how to choose the latent parameters of this generative model in order to find a reasonable prior.
\subsection{The likelihood}\label{subsec:likelihood}
The likelihood is the conditional probability of a data realization $d$ given the underlying physical signal $s$. 
In the case of photon-count instruments like eROSITA, this conditional probability for a pixel $i$, takes the form of a Poisson distribution
\begin{equation}
    \mathcal{P}(d_i|s) = \mathcal{P}(d_i|\lambda_i(s)) = \frac{\lambda_i^{d_i}}{d_i!}e^{-\lambda_i} \ ,\label{eq:likelihood_prob_index}
\end{equation}
with $d_i$ being the photon counts and $\lambda_i$ being the mean photon flux on the detector pixel $i$, caused by the signal $s$. 
For a CCD chip with $n$ instrument pixels, the data is a vector of pixel photon counts, $d = (d_i)_{i \in \{1,...n\}}$. 
The total likelihood turns into the product of the individual likelihoods in the case of statistical independence of the pixel events, 
\begin{equation}
    \mathcal{P}(d|s) = \prod_i \mathcal{P}(d_i|\lambda_i(s)) = \prod_i \frac{\lambda_i^{d_i}}{d_i!}e^{-\lambda_i}\label{eq:likelihood_prob} \ .
\end{equation}
Often we refer to the negative logarithm of this probability as the information Hamiltonian of the likelihood
\begin{equation}
     \mathcal{H}(d|s) =  - \ln \mathcal{P}(d|\lambda(s)) =   \sum_i \lambda_i - d_i \ln(\lambda_i) + \ln(d_i!) \label{eq:likelihood_ham} \ .
\end{equation}
These equations can be generalized to multiple observations $m$ of the same sky with different instruments or at different times.
Then the data is a vector of vectors, $d = (d_j)_{j \in \{1,...m\}}=(d_{ji})_{j \in \{1,...m\}, \  i \in \{1,...n\}}$, where $d_{ji}$ is the data point from the pixel $i$ in the observation $j$ which turns Eq.~(\ref{eq:likelihood_ham}) into
\begin{equation}\label{eq:full_likelihood_ham}
     \mathcal{H}(d|s) =  \sum_j \mathcal{H}(d_j|\lambda_j(s)) =  \sum_{j, i} \lambda_{ji} - d_{ji} \ln(\lambda_{ji}) + \ln(d_{ji}!) \ .
\end{equation}
The steps performed to bin the data before using it in this formula are explained in Sect.~\ref{sec:data}.
In order to evaluate the Hamiltonian $\mathcal{H}(d|s)$  we need a digital representation of the measurement process, the relation between the physical signal $s$, and the expected number of counts $\lambda$.
The derivation of these quantities is discussed in the following section (Sect.~\ref{subsec:instrument_model}).

\subsubsection{Instrument model}\label{subsec:instrument_model}
An accurate description of the measurement process is essential for the inference of the signal $s$.
Therefore, we need the instrument response $R$, which represents the effects of the measurement process,
\begin{equation}
  \label{eq:response}
\lambda = R(s),
\end{equation}
to be as accurate as possible.
However, since this function will be called many times during the computation process, it also has to be efficient and therefore we aim for a representation that is not only precise but also computationally affordable.
In essence, we want to build a forward model that describes the linear effects of the measurement process. We tackle this by subdividing the response function $R$ into its most relevant constituents.
The photon flux $s$ coming from the sky gets smeared out by the \ac{PSF} of the \ac{MA}. This gets mathematically described by operator $O$.
The \ac{PSF} of each individual mirror module on-ground, on-axis and in-focus is of the order of $16.1 \arcsec$. 
However, the modules are mounted intra-focal to reduce the off-axis blurring for the price of an enlarged \ac{PSF} in the on-axis region. 
Therefore, the in-flight on-axis \ac{PSF} is approximately $18 \arcsec$, and the averaged angular resolution of the field of view is improved to approximately $26 \arcsec$~\citep{Predehl2020}.
The blurred flux gets then collected by the \ac{CA}. 
We denote the mathematical operator representing the exposure with $E$. It encodes the observation time and detector sensitivity effects. The flagging of invalid detector pixels, also called the mask, is denoted with $M$. 
The instrument response is thus
\begin{equation}
  \label{eq:detailed_response_continuos}
R = M \circ E \circ O\, ,
\end{equation}
where $\circ$ denotes the composition of operators.
Readout streaks are almost completely suppressed due to the fast shift from the imaging to the frame-store area of the \acp{CCD} and therefore, don't have to be modeled \citep{Predehl2020}.
Other effects, like pile-up are neglected up to this moment, but will be covered in future work. 
In the following sections, the parts of the instrument response will be discussed individually.
\subsubsection{The point spread function}
The \ac{PSF}, here denoted as the mathematical operator $O$, describes the response of the instrument to a point-like source. An incoming photon from direction $x \in \mathbb{R}^2$ is deflected to a different direction $\Tilde{x} \in \mathbb{R}^2$. 
This blurs the original incident flux $s$ to the blurred flux $s^\prime$, which is notated in a continuous and discretized way,
\begin{equation}\label{eq:fredholm}
    s^\prime(\Tilde{x}) = \int_{\mathbb{R}^2} O(\Tilde{x},x)\, s(x) \dd{x} \\
    s^\prime_{\Tilde{x}} = \sum_{x} O_{\Tilde{x}x}\,  s_{x} \ .
\end{equation}
This operator $O(\Tilde{x},x)$ can be regarded as a probability density function $\mathcal{P}(\Tilde{x}|x)$, which is normalized by the integration over the space of all directions meaning, that the process of blurring conserves the photon flux, 
\begin{equation}
    1  = \int_{\mathbb{R}^2} O(\Tilde{x},x) \dd{\Tilde{x}} = \int_{\mathbb{R}^2} \mathcal{P}(\Tilde{x}|x) \dd{\Tilde{x}} \ .
\end{equation}

In the discretized form the operator $O$ is a matrix and thus scales quadratically with the number of pixels $n$, resulting in a computational complexity of $\mathcal{O}(n^2)$. 
In most applications, a spatially invariant \ac{PSF} is assumed, meaning that the \ac{PSF} is the same for all points within the field of view. Thus the \ac{PSF} is only a function of the deflection $\Tilde{x}-x$, meaning $O(\Tilde{x}, x) = O(\Tilde{x}-x)$. This fact turns Eq.~(\ref{eq:fredholm}) into a convolution
\begin{equation}\label{eq:convolution}
    s^\prime(\Tilde{x}) = \int_{\mathbb{R}^2} \mathrm{O}(\Tilde{x}-x)\, s(x) \dd{x}.
\end{equation}
Convolutions on regular grids can be executed very efficiently, thanks to the convolution theorem and the \ac{FFT} developed by \cite{cooley1965algorithm}.
However, the assumption of spatial invariance of the \ac{PSF} only holds, depending on the variability of the \ac{PSF}, for smaller fields of view.
To image large structures in the sky, the spatial variability of the \ac{PSF} cannot be ignored without imprinting artifacts on the reconstructions.

Therefore, we need a representation of spatially variant \acp{PSF} that can be used in the forward model.
Here, we use the algorithm of \cite{nagy1997fast}. This algorithm, which we call \textit{linear patched convolution} in the following, is a method to approximate spatially variant \acp{PSF} in a computationally efficient way.
It scales sub-quadratically, meaning it is computationally afforable, but improves the accuracy, in comparison to a regular spatially in-variant convolution.

In \textit{linear patched convolution} the full spatially-variant \ac{PSF}, $O$, is approximated by a combination of operations
\begin{equation}\label{eq:approx_psf}
    O  \approx \sum_k P_k W C_k \ .
\end{equation}
First, the image is cut into $k$ overlapping patches by the slicing operator $C_k$. Next, these patches are weighted with a linear interpolation kernel $W$, such that the total flux $s$, despite the overlapping patches, is conserved. Then, each patch is convolved with the associated \ac{PSF} corresponding to the center of the patch, denoted by $P_k$. Finally, the results of the weighted and convolved overlapping patches are summed up. This can be seen as an Overlap-Add convolution with linear interpolation and different \acp{PSF} for each patch (see \cite{nagy1997fast}).

In order to perform this operation, we need information about the spatial variability  of the \ac{PSF} across the \ac{FOV}, which we can retrieve from the \ac{CalDB}\footnote{Information about the \ac{CalDB}: \url{https://erosita.mpe.mpg.de/edr/DataAnalysis/esasscaldb.html}}. Here, we find information about the \ac{PSF}, gathered at the PANTER $\SI{130}{meter}$ long-beam X-ray experimental facility of the Max-Planck-Institute for Extraterrestrial Physics \citep{Predehl2020}.\footnote{Details in Appendix \ref{app:data}} 
The \ac{CalDB} files contain the measurements of the \ac{PSF} for certain off-axis angles and energies, averaged over the azimuth angle. 

\begin{figure}[!h]
  \centering
  \includegraphics[width=\linewidth]{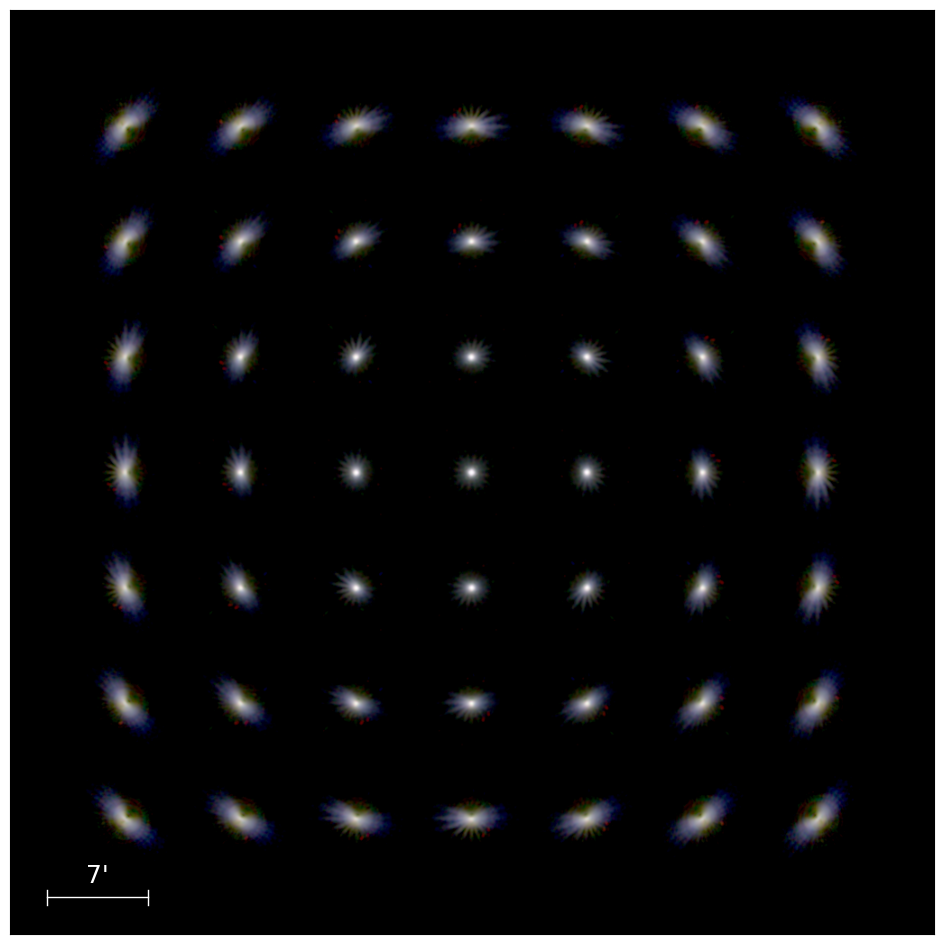}
  \caption{Visualization of eROSITA PSF approximated by the \textit{linear patch convolution} algorithm for three energy bins. The different colors represent the logarithmic intensities in the three energy bands. Red: 0.2-1.0 keV, Green: 1.0-2.0 keV, Blue: 2.0-4.5 keV.}
   \label{fig:psf}
\end{figure}

For the \textit{linear patched convolution} algorithm at use we need the \ac{PSF} at the central positions in the patches.
To obtain these, we rotate and linearly interpolate the \acp{PSF} from the \ac{CalDB}, which allows us to construct the PSFs at these central positions.
We also remove some noticeable shot noise from the measured \acp{PSF} by clipping the normalized \acp{PSF} at $10^{-6}$.
A visualization of the approximated \ac{PSF} of eROSITA can be seen in Fig.~\ref{fig:psf}. More detailed information on the eROSITA PSF can be found in \cite{Predehl2020}.

\subsubsection{The exposure}
The received flux $\lambda$ on the camera is observed for a total exposure time $\Theta$ by the CA.
The exposure operator $E$ includes not only the exposure time $\Theta$, but also the vignetting, $\rho$,  of the \ac{TM}, and its effective area $\mu$.
In the case of a time-invariant flux $s^\prime(t) = s^\prime_0$, the integral over time corresponds to a multiplication with the total exposure time $\Theta$ and thus
\begin{equation}
    \lambda = \int_\Theta \rho \mu s^\prime(t) \dd{t} =  \rho \mu s^\prime_0\int_\Theta \dd{t} = (\rho \mu \Theta) s^\prime_0=  E s^\prime_0.
\end{equation}
In the case that $s^\prime(t)$ is not constant, $s^\prime_0$ is the average value of $s^\prime(t)$ in the observed time interval. 
We calculate the total observation time of a pixel projected to the sky in a certain energy band, combined with the vignetting, with \texttt{eSASS} (eROSITA Science Analysis Software System)\footnote{More information about the \texttt{eSASS} software developed by the eROSITA Team can be found here: \url{https://erosita.mpe.mpg.de/edr/DataAnalysis/}}\citep{Brunner2018, Predehl2020}
through the command \texttt{expmap}. The parameters used for the \texttt{expmap} command can be found in ~\ref{app:data} in Table~\ref{tab:expmapparams}. The information about the effective area for each \ac{TM} can be found in the \ac{CalDB}\footnote{Details in Appendix~\ref{app:data}}.

\subsubsection{The mask}

The information Hamiltonian of the likelihood, Eq.~(\ref{eq:full_likelihood_ham}), derived from the Poisson distribution, is only defined for $\lambda > 0$, as it includes a logarithmic term in the count rate $\lambda$. Consequently, it is necessary to mask all sky positions with zero exposure time or defective detector pixels for a given \ac{TM}, as these would result in $\lambda = 0$ and thus violate the assumptions of the likelihood model.

Removing these pixels from the calculation makes the algorithm more stable, prevents the appearance of \texttt{NaNs}, and ensures that only reliable data is used for the reconstruction. From the raw data, there seem to be corrupted data points in regions with very low exposure time and at the boundary of the \ac{FOV}. 
Therefore, we decided to mask from the reconstruction all pixels and data points with an observation time of less than $500$~seconds.

Since not all bad pixels were correctly flagged in the exposure files, we used information from the \ac{CalDB} badpix files to update the detmap files. We then used these modified detmap files to build the new exposure maps and update the mask.

\subsubsection{Forward model for multiple observations}
\begin{figure}[!h]
  \centering
  \begin{tikzpicture}[scale=0.8]
%\tikzstyle{i}= [circle,draw, minimum size=0.8cm,inner sep=4pt, outer sep=0pt]
%\begin{scope}[shift={(0,2)}]
\node at (0, 1) [draw, rectangle] (s){\textsc{Signal}};
\node at (0, 0) [draw, rectangle] (tm3){$\mathcal{R}_{3}$};
\node at (-1, 0) [draw, rectangle] (tm2){$\mathcal{R}_{2}$};
\node at (-2, 0) [draw, rectangle] (tm1){$\mathcal{R}_{1}$};
\node at (1 ,0) [draw, rectangle] (tm4){$\mathcal{R}_{4}$};
\node at (2, 0 ) [draw, rectangle] (tm6){$\mathcal{R}_{6}$};
\node at (0, -1) [draw, rectangle] (l3){$\lambda_3$};
\node at (-1.5, -1) [draw, rectangle] (l2){$\lambda_2$};
\node at (-3, -1) [draw, rectangle] (l1){$\lambda_1$};
\node at (1.5, -1) [draw, rectangle] (l4){$\lambda_4$};
\node at (3, -1)[draw, rectangle] (l6){$\lambda_6$};
\node at (0, -2) [draw, rectangle] (h3){$\mathcal{H}(d_{3}|\lambda_{3})$};
\node at (-2, -2)[draw, rectangle] (h2){$\mathcal{H}(d_{2}|\lambda_{2})$};
\node at (-4, -2)[draw, rectangle] (h1){$\mathcal{H}(d_{1}|\lambda_{1})$};
\node at (2, -2) [draw, rectangle] (h4){$\mathcal{H}(d_{4}|\lambda_{4})$};
\node at (4, -2)[draw, rectangle] (h6){$\mathcal{H}(d_{6}|\lambda_{6})$};
\node at (0, -3.5)[draw, rectangle] (h_all){$\mathcal{H}(d|s)$};
\draw[->] (s) to (tm1);
\draw[->] (s) to (tm2);
\draw[->] (s) to (tm3);
\draw[->] (s) to (tm4);
\draw[->] (s) to (tm6);
\draw[->] (tm1) to (l1);
\draw[->] (tm2) to (l2);
\draw[->] (tm3) to (l3);
\draw[->] (tm4) to (l4);
\draw[->] (tm6) to (l6);
\draw[->] (l1) to (h1);
\draw[->] (l2) to (h2);
\draw[->] (l3) to (h3);
\draw[->] (l4) to (h4);
\draw[->] (l6) to (h6);
\draw[->] (h1) to (h_all);
\draw[->] (h2) to (h_all);
\draw[->] (h3) to (h_all);
\draw[->] (h4) to (h_all);
\draw[->] (h6) to (h_all);
%\end{scope}
\end{tikzpicture}
  \caption{Visualization of the computational graph of the forward model.}
   \label{fig:comp_graph}
\end{figure}
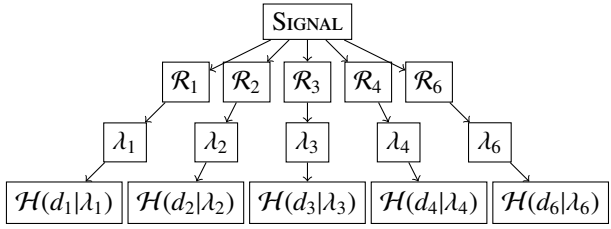

In the case of an eROSITA pointing, all \acp{TM} that are online observe the same sky and capture the same physics.
Although the instruments are very similar, they are not identical. 
Their slightly different pointing results in different positions of the focal point of the \ac{PSF}. 
Also, they may have different good-time intervals, resulting in different exposure times and also \ different defective pixels for the \acp{CCD}.
Instead of summing the counts from the different data sets, thereby assuming a ``mean'' instrument, we model each \ac{TM} and its observation individually. 
That means, we formulate the signal response $\lambda_j$ of one $\mathrm{TM_j}$ as 
\begin{equation}
  \label{eq:detailed_response_index}
\lambda_j = \mathrm{M_jE_jO_j}s \ .
\end{equation}
We display a visualization of the forward model's computational graph in Fig.~\ref{fig:comp_graph}. By plugging in all $\lambda_j$ into Eq.~\ref{eq:full_likelihood_ham}, we get a formulation of the full likelihood information Hamiltonian that allows us to remove the individual detector effects jointly.

\subsection{Inference}\label{subsec:inference}

In principle, given the prior and likelihood distributions, Eq.~(\ref{eq:bayes}) allows to fully determine the posterior distribution by computing the evidence
\begin{align}
	\mathcal{P}(d) = \int_{\Omega_s} \mathcal{P}(d|s)\, \mathcal{P}(s)\, \mathcal{D}s,
\end{align}
where we have denoted with $\Omega_s$ the Hilbert space in which $s$ lives.
In general, and specifically for the prior and likelihood models described above, the evidence cannot be explicitly evaluated, as it would require integrating over the potentially multi-million- or multi-billion-dimensional space, $\Omega_s$.
To overcome this problem we use \ac{VI}. In \ac{VI}, the evidence calculation problem is overcome by approximating the posterior distribution directly using a family of tractable distributions $\mathcal{Q}_\alpha (s|d)$, parametrized by some variational parameters $\alpha$.
To approximate the posterior we minimize the Kullback-Leibler divergence
\begin{align}
	\mathcal{D}_\text{KL}\qty(\mathcal{Q_\alpha}||\mathcal{P}) \coloneqq \int_{\Omega_s} \mathcal{Q}_\alpha\qty(s|d)\, \log{\frac{\mathcal{Q}_\alpha\qty(s|d)}{\mathcal{P}\qty(s|d)}}\, \mathcal{D}s,
\end{align}
with respect to the variational parameters $\alpha$.
In this work, the family of approximating posterior distributions $\mathcal{Q}_\alpha(s|d)$ is built using geometric VI (geoVI, \cite{geovi}).
In geoVI, the posterior is approximated with a Gaussian distribution in a space in which the posterior is approximately Gaussian.
This is achieved by utilizing the Fisher information metric, which captures the curvature of the likelihood and the prior distributions. The Fisher metric provides a way to measure the local geometry of the posterior, guiding the creation of a local isometry -- a transformation that maps the curved parameter space to a Euclidean space while preserving its geometric properties.
In this transformed space, the posterior distribution approximates a Gaussian distribution more closely, allowing the Gaussian variational approximation to be more accurate. Consequently, geoVI can represent non-Gaussian posteriors with high fidelity, improving inference results.
By leveraging the geometric properties of the posterior distribution, geoVI offers a powerful extension to traditional VI, enabling more precise and reliable approximations for complex Bayesian models, as the ones presented in this work.

\section{Results}\label{sec:results}

\begin{figure*}[!h]
  \centering
  \includegraphics[width=0.49\linewidth]{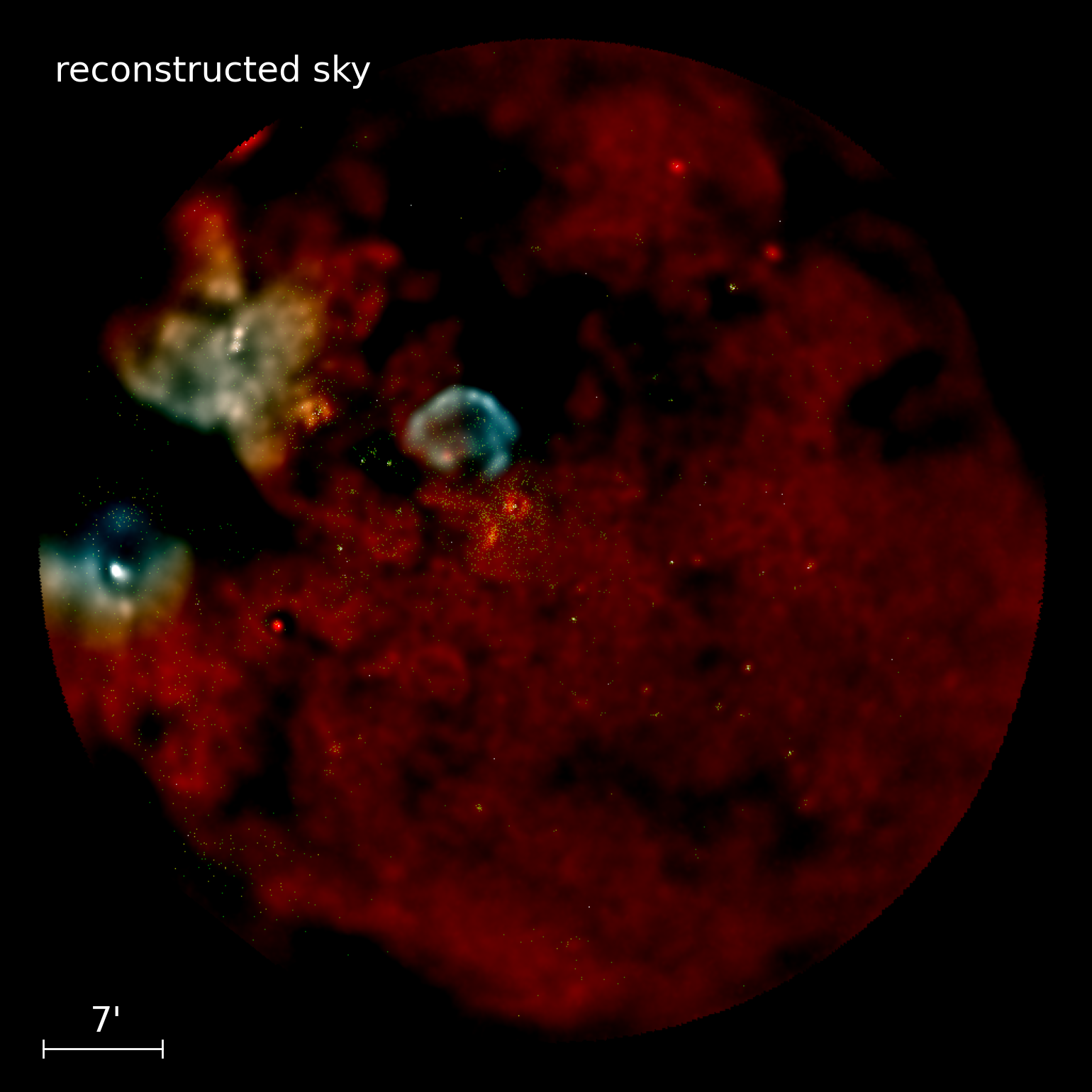}
  \includegraphics[width=0.49\linewidth]{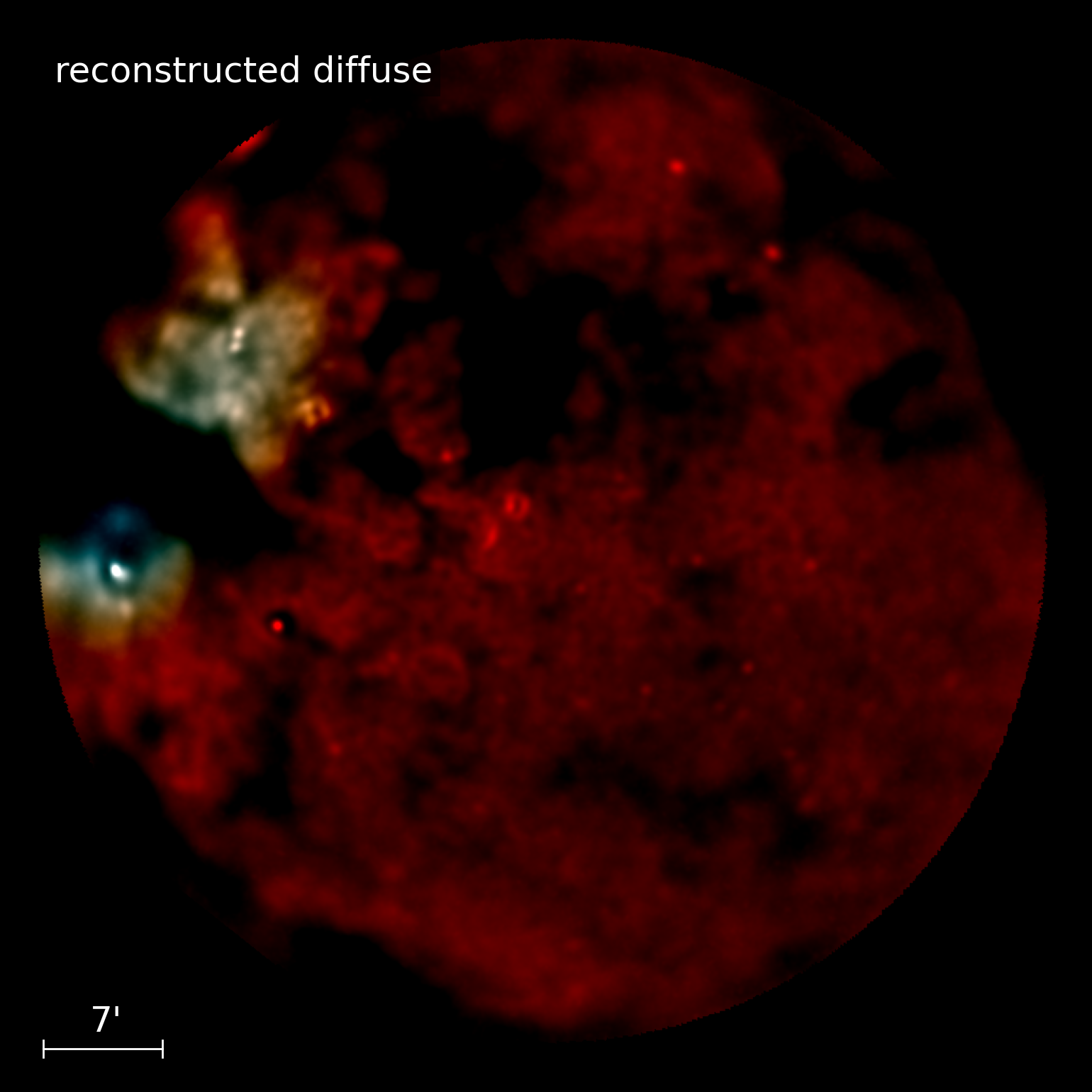} \\
  \includegraphics[width=0.49\linewidth]{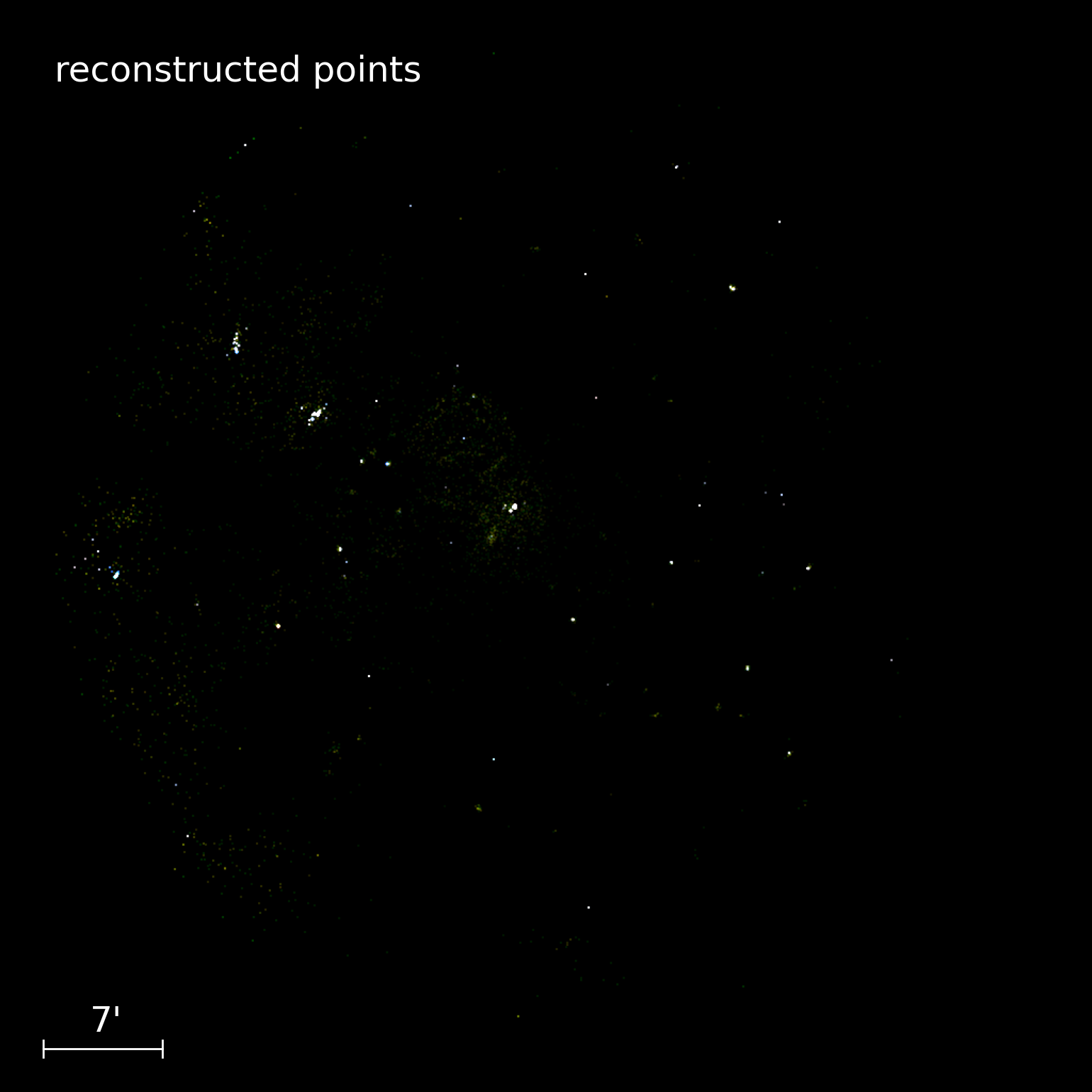}
  \includegraphics[width=0.49\linewidth]{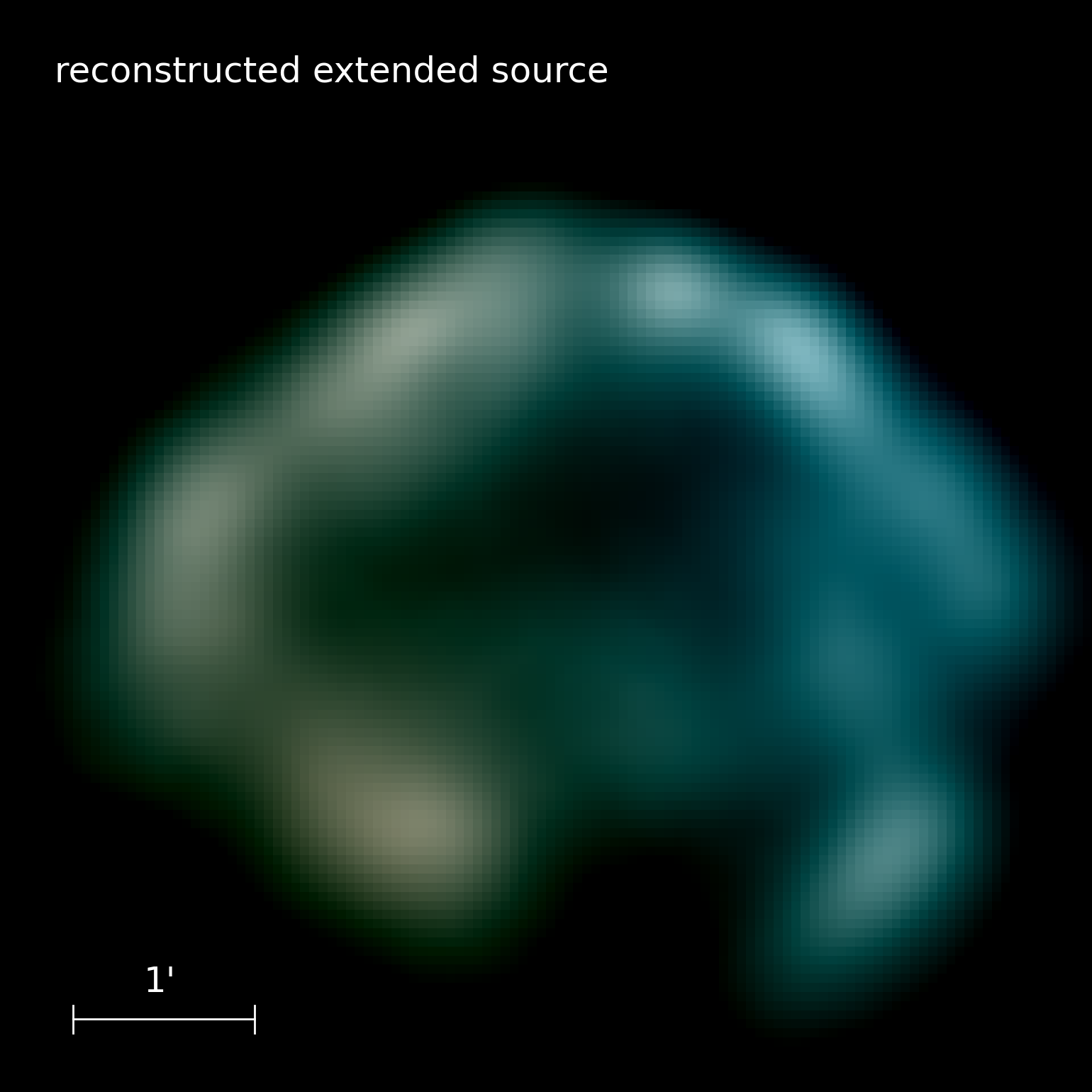}
  \caption{
  Posterior mean of the SN1987A reconstruction. 
  \textit{The top panels} display on the left the reconstruction of the sky and on the right the separated diffuse emission.
  \textit{The bottom panels} display the reconstruction of the point-like emission (left) and a zoom into the reconstruction of the diffuse emission from 30 Doradus C (right) as marked in Fig.~\ref{fig:Data}. 
  We convolve the point sources with an unnormalized Gaussian kernel with standard deviation $\sigma=0.5$, in order to make them visible on printed paper.
  The different colors represent the logarithmic intensities in the three energy channels $0.2\ \text{-}\ 0.1\; \si{\kilo\electronvolt}$, $1.0\ \text{-}\ 2.0\; \si{\kilo\electronvolt}$, and $2.0\ \text{-}\ 4.5\; \si{\kilo\electronvolt}$ and are depicted in red, green, and blue, respectively.
  }
  \label{fig:full_rec}
\end{figure*}

In Figure~\ref{fig:full_rec}, we present the reconstruction of the sky flux distribution based on the data shown in Figure~\ref{fig:Data}. Our algorithm’s forward modeling of the X-ray sky enables the decomposition of the signal into point-like, diffuse, and extended-source emission components, providing a more detailed view of the small-scale features of the extended structure of the 30 Doradus C bubble. These reconstructed components are also displayed in Figure~\ref{fig:full_rec}. From these reconstructions, it is clear that most of the point-like emission is well separated from the diffuse emission, resulting in the first denoised and deconvolved view of this region of the sky as observed by the eROSITA X-ray observatory. 
Additionally, in Fig.~\ref{fig:rec_per_energybin}, we show the reconstructed flux for each energy bin, offering a clearer understanding of the color scheme adopted in Fig.~\ref {fig:full_rec}.
All the final reconstructions have been obtained using the geoVI algorithm.
For the spatial distribution, we have chosen a resolution of $1024\times1024$ pixels.
For the spectral distribution, we have chosen $3$ energy bins corresponding to the energy ranges between $0.2\ \text{-}\ 0.1\; \si{\kilo\electronvolt}$, $1.0\ \text{-}\ 2.0\; \si{\kilo\electronvolt}$, and $2.0\ \text{-}\ 4.5\; \si{\kilo\electronvolt}$, respectively.
The variational approximation to the posterior was estimated using $8$ samples, corresponding to $4$ pairs of antithetic samples. We considered the posterior approximation converged when the posterior expectation values of the signals of interest, such as the reconstructed sky flux field, exhibited no significant changes between consecutive iterations of the \ac{VI} algorithm.
Specifically, we considered the algorithm converged after at least three consecutive geoVI iterations during which the mean squared weighted deviations remained below $1.05$.
The runtime for the reconstruction was approximately one day on a CPU for a single module, and around two days for all five analyzed telescope modules.
By adopting a fully probabilistic approach, we leverage posterior samples to assess how well the model assumptions align with the observed data. 
In particular, in the presence of shot noise, we define the posterior mean of the \ac{NWR} as
\begin{align}
    r_\text{NWR} = \bigg \langle \frac{\lambda(s) - d}{\sqrt{\lambda(s)}} \bigg \rangle_{\mathcal{Q}_\alpha(s|d)} \simeq \frac{1}{N_s} 
    \sum_{i=0}^{N_s} \frac{\lambda(s^*_i) - d}{\sqrt{\lambda(s^*_i)}}, \label{eq:nwr}
\end{align}
where $\lambda(s)$ is the expected number of counts predicted by the model, $d$ is the observed data, and $N_s$ is the number of approximate posterior samples.
Here, the posterior average over $\mathcal{Q_\alpha}$  is approximated by the sample average over the corresponding posterior samples $s^*_i$.
These residuals are particularly useful for identifying model inconsistencies, which may indicate areas for improving the instrument’s description as well as point to potential calibration improvements. We will explore this possibility further in the following section (Sect.~\ref{sec:discussion}).

\section{Validation and discussion}
\label{sec:discussion}
The discussion is divided into two parts. First, in Sect.~\ref{sec:validation}, we validate the general consistency of the presented algorithm using a simulated sky and simulated data, which also motivates the detection threshold for point sources. In the second part, Sect.~\ref{sec:resultsdiscussion}, we discuss the results of the reconstruction presented in Sect.~\ref{sec:results}, along with corresponding diagnostics, such as the \acp{NWR}.
\subsection{Validation}
\label{sec:validation}
\begin{figure*}[!h]
  \centering
  \includegraphics[width=0.33\linewidth]{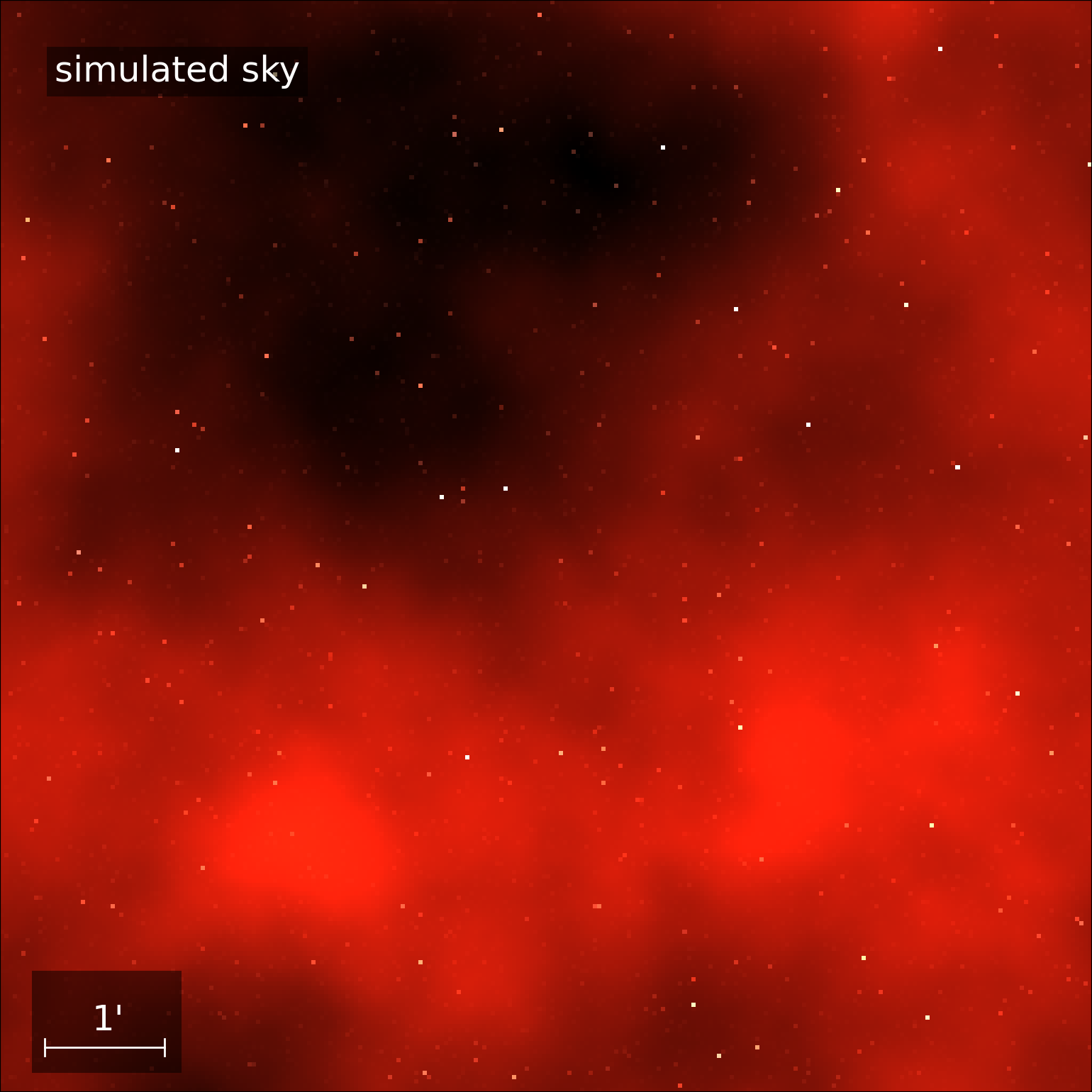} 
  \includegraphics[width=0.33\linewidth]{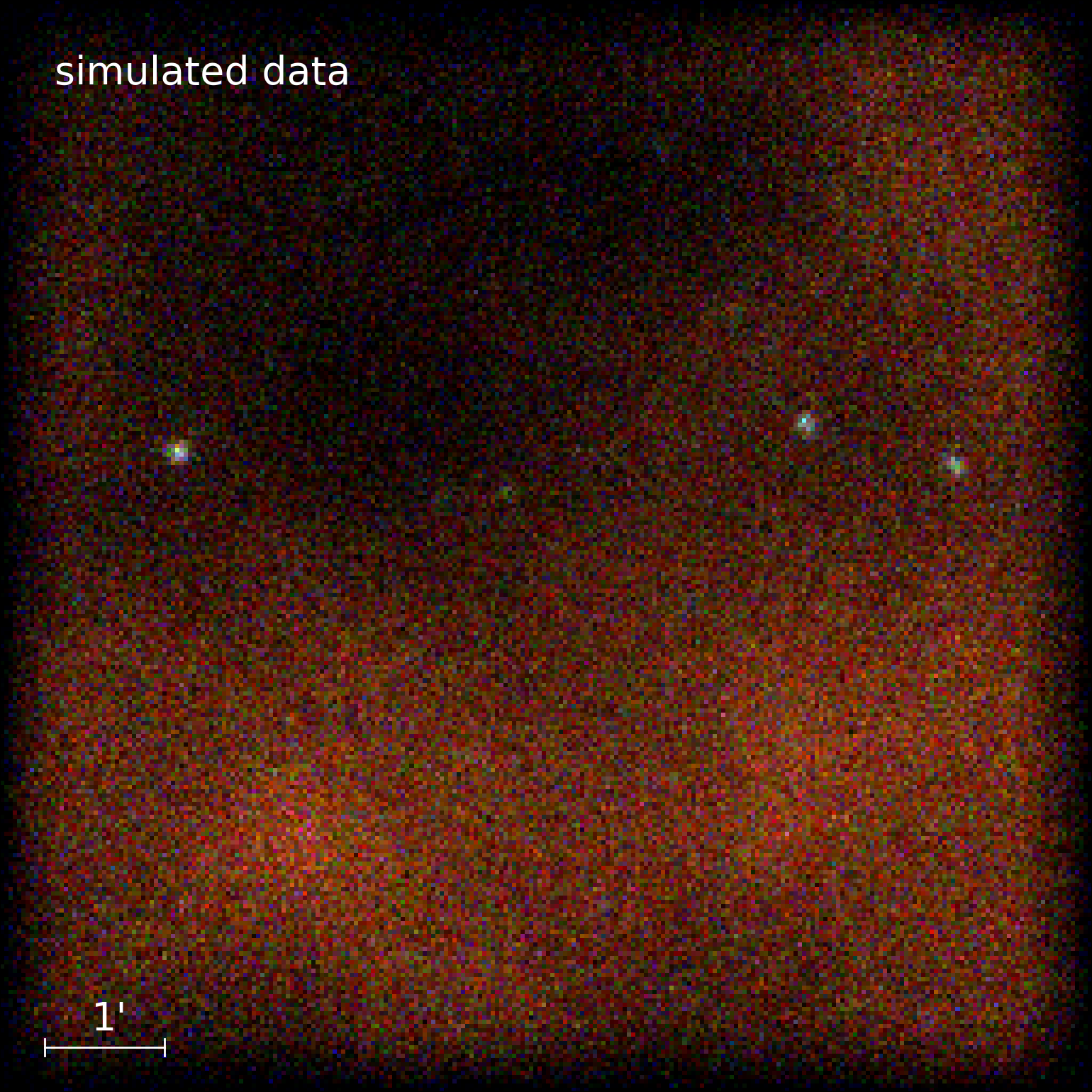} 
  \includegraphics[width=0.33\linewidth]{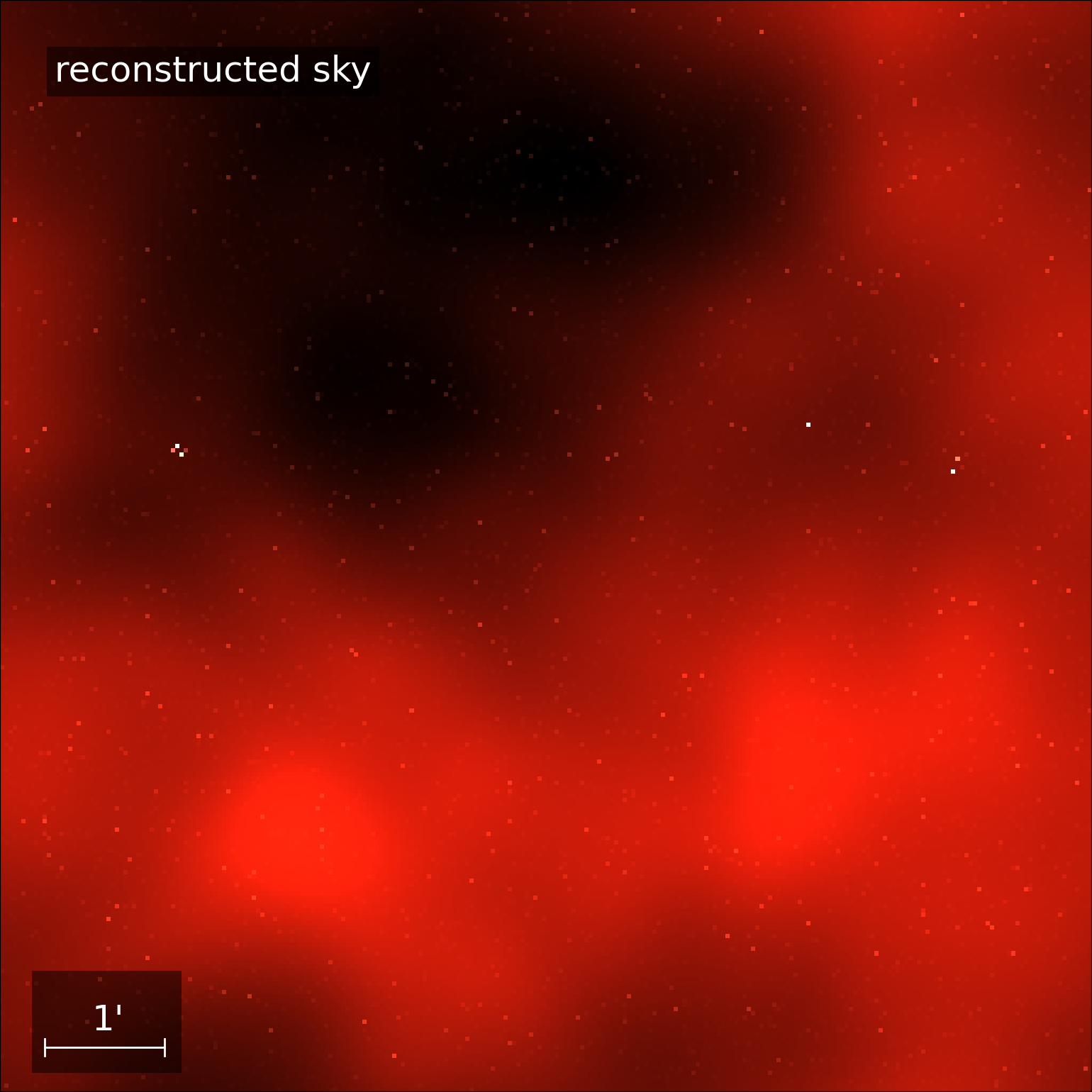} 
  \caption{Visualization of the validation of the imaging algorithm. Left: Simulated X-ray sky. Center: Simulated X-ray data generated as shown in Fig.~\ref{fig:comp_graph}. Right: Reconstructed X-ray sky.}
  \label{fig:validation}
\end{figure*}

Generative modeling allows to generate prior models of the sky, as described in Sect.~\ref{subsec:prior_models}. 
These prior models can be used to validate the consistency of the presented algorithm. 
In particular, we look at prior samples of the X-ray sky, composed of point sources and diffuse emission, with a FOV of $1024$ arcsec. Using the same resolution as for the actual reconstruction, this leads to $256 \times 256$ pixels. We pass the prior samples through the forward model shown in Fig.~\ref{fig:comp_graph}, including all five \acp{TM}, which gives us simulated data. Figure~\ref{fig:validation} shows the considered prior sample of the X-ray sky as well as the corresponding simulated data passed through the eROSITA response and affected by Poissonian noise. The simulated data per \ac{TM} and the underlying simulated sky per energy bin is shown in the Appendix~\ref{app:validationdiagnostics} in Figs.~\ref{fig:simdata_ebin} and \ref{fig:simsky_ebin}. Using the simulated data $\tilde{d}$, we aim to apply the algorithm presented above to estimate the posterior $\mathcal{P}(s|\tilde{d})$ through \ac{VI} posterior samples. We will then evaluate how well the corresponding prior $\mathcal{P}(s)$ is reconstructed and determine the corresponding uncertainty in that estimate. The right side of Fig.~\ref{fig:validation} shows the reconstructed prior sample. Component separation, deconvolution, and denoising techniques show strong performance when applied to simulated data. They effectively recover the underlying signal. \\
To validate the results shown in Fig.~\ref{fig:validation}, we use a set of validation metrics that we have access to due to the probabilistic approach of the algorithm. These metrics are intended to provide further insight into the residuals between the simulated X-ray sky and its reconstruction, as well as the uncertainty of the algorithm at each pixel. 

\begin{figure}[!h]
  \centering
  \includegraphics[width=\linewidth]{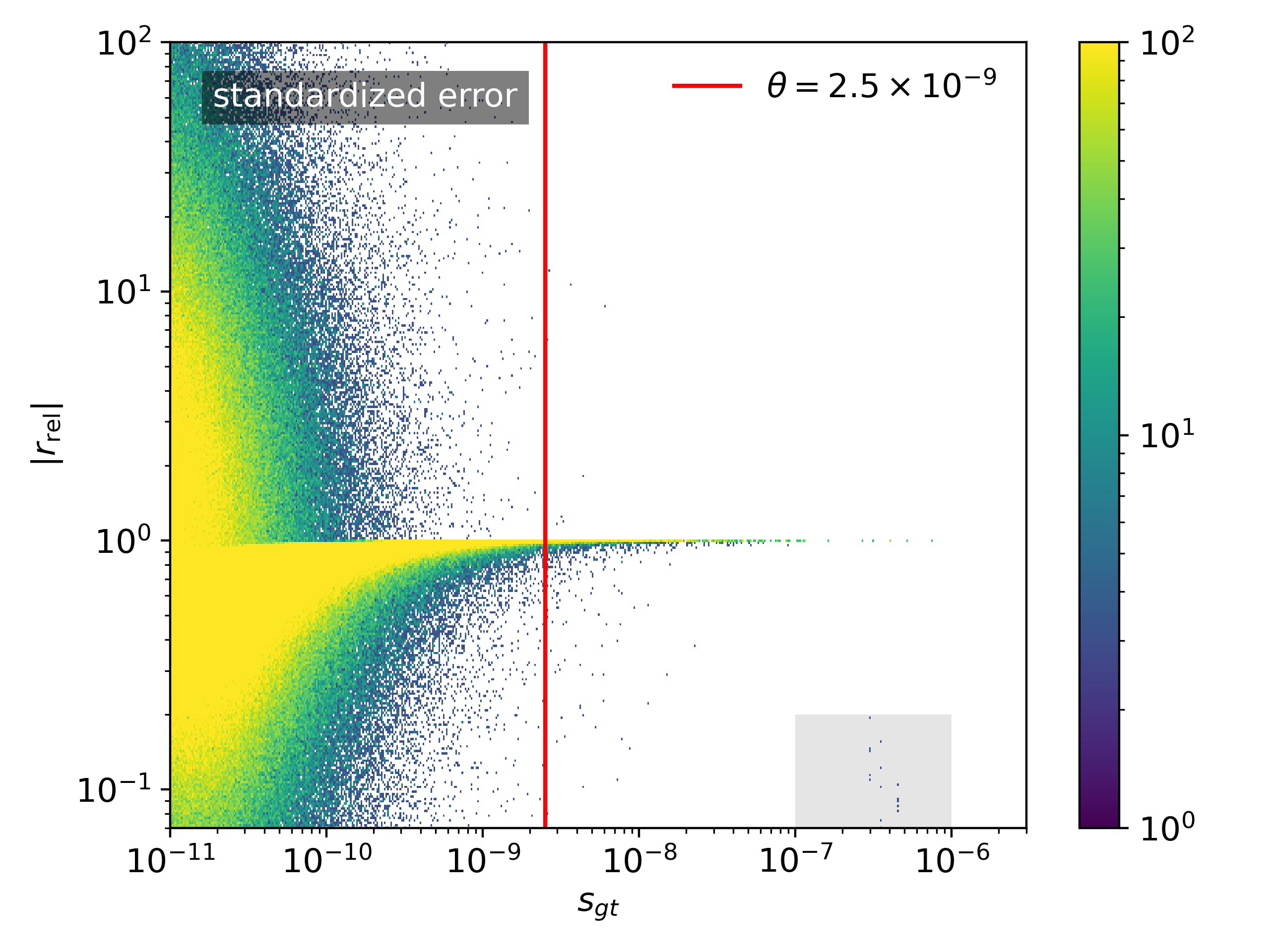} 
  \caption{Two-dimensional histogram of the standardized error (Eq.~\ref{eq:standerror_abs}) for the point sources. The histogram is plotted together with the lowest detection threshold, $\theta = 2.5 \times 10^{-9}$, calculated for the region with the longest observation time in Appendix~\ref{app:hyperparams}. The colorbar shows the counts per bin in the two-dimensional histogram, i.e., higher values correspond to more frequent combinations of standardized error and ground truth flux.}
  \label{fig:2dhist}
\end{figure}

Accordingly, we show in the appendix the standard deviation of the posterior samples in Fig.~\ref{fig:simulateduncertainty}, which gives us a measure of the uncertainty of the algorithm. To examine the residuals, we define the standardized error as the relative residual between the ground truth, $s_\text{gt}$, and the posterior mean, $s$,
\begin{align}
    r_{\text{rel}}(s_\text{gt}) = \frac{s-s_{\text{gt}}}{s_\text{gt}},
    \label{eq:standerror_sign}
\end{align}
to check for differences between the ground truth and the reconstruction. This standardized error is shown for each energy bin in Appendix \ref{app:validationdiagnostics} in Fig.~\ref{fig:standardized error}. The image shows that point sources are not detected or are misplaced in some areas. This highlights the need for a detection threshold for point sources in the reconstruction to ensure the correctness as also indicated in the hyper parameter search in Appendix~\ref{app:hyperparams}. To validate the detection threshold further we use posterior samples, $s_p^*$,  of the approximated posterior, $Q(s|d)$, for the point source component in order to get the absolute sample-averaged two-dimensional histogram of the standardized error only for point sources, $|r^*_{\text{rel}}(s_\text{gt, p})|$ , where, 
\begin{align}
    |r^*_{\text{rel}}(s_\text{gt, p})| =  \bigg \vert\frac{s_\text{p}^*-s_{\text{gt, p}}}{s_\text{gt, p}} \bigg \vert.
    \label{eq:standerror_abs}
\end{align}
Figure~\ref{fig:2dhist} shows the sample-averaged histogram together with the detection threshold, $\theta$, analytically set for this reconstruction in the Appendix~\ref{app:hyperparams}. The figure shows that, even above the threshold, the standardized error for many point sources remains close to 1. These sources are relevant but not identified in the reconstruction due to noise and instrumental effects. This is also evident in Figure~\ref{fig:validation}, which compares the ground truth to the reconstructed signal. Importantly, this behavior is still acceptable: the purpose of the threshold is not to guarantee detection of all real sources, but rather to control false detections, i.e., the appearance of point sources where none exist. As the diagram indicates, such spurious detections may occur below the threshold. Point sources that are clearly identified are highlighted in the gray box in Figure~\ref{fig:2dhist}. This interpretation is further supported by Figure~\ref{fig:standardized error} in the appendix. In summary, below the detection threshold, the histogram shows two effects, undetected or misplaced point sources and possible noise over-fitting, which are eliminated by cutting the point sources below the detection threshold to ensure the consistency of the reconstruction. We apply the same cuts to the reconstruction shown in Sect.~\ref{sec:results}.
We note that the threshold on the reconstructed point-source field is applied only a posteriori to isolate emission from reliably detected point sources. Consequently, unresolved point sources remain statistically distributed between the diffuse emission component and the underlying inverse-Gamma distributed field of faint, unresolved point sources. This statistical separation is further guided by the inferred spatial and spectral correlation kernel of the diffuse emission: since spatial correlations are learned during inference, flux from unresolved point sources is more likely to be correctly assigned to the true diffuse component. However, while this holds true for synthetic observations, additional extinction processes in real observational data render this separation even less constrained.

\subsection{Discussion of results}
\label{sec:resultsdiscussion}
The results of the algorithm described above, applied to the eROSITA LMC data, are shown in Sect.~\ref{sec:results}. Fig.~\ref{fig:full_rec} shows the LMC in a deconvolved, denoised and decomposed view. The full image of the \ac{LMC} is shown, as well as the separated components of the point sources, the diffuse structures of the \ac{LMC}, and the extended sources of 30 Doradus C. As a result of the inference, we get posterior samples of the approximated posterior probability $Q(s|d)$. Given these posterior samples, we can calculate a measure of uncertainty of the reconstruction, which is in this case given by the standard deviation. The corresponding plots of the standard deviation per energy bin for the reconstruction shown in Fig.~\ref{fig:full_rec} is shown in Appendix~\ref{app:resultsdiag} in Fig.~\ref{fig:rec_unc_per_energybin}. As expected, we can see that the uncertainty is higher in regions with a high number of photon counts. This reflects the fact that in regions of higher flux the uncertainty is also higher.\\
Analyzing the component separation in Fig.~\ref{fig:full_rec}, it can be seen that there is still a halo around the central SN1987A source, which can have two different causes. First, it could be due to a detection pile-up effect caused by the high fluxes from these sources \citep{Davis_2001}. Second, it could be due to a mismodeling of the instruments caused by calibration mismatches. In order to check for possible calibration issues, we performed single-\ac{TM} reconstructions, which only took the data and the response functions for one of the \acp{TM} each into account. The results of the single-\ac{TM} reconstructions per energy bin are shown in Appendix~\ref{app:resultsdiag} in Fig.~\ref{fig:rec_tm1}. These images give us a great insight into possible calibration inconsistencies together with the corresponding \acp{NWR} (Eq.~\ref{eq:nwr}) per \ac{TM} and energy bin, which are shown in Fig.~\ref{fig:nwr_tm_1} and Fig.~\ref{fig:nwr_tm_2}. In particular, the reconstruction for \ac{TM}2 suggests both pile-up issues and mismatches in the calibration files, such as the \ac{PSF} and dead pixels. Although we incorporated information about the dead pixels into the inference, the number of dead pixels accounted for seems to be insufficient. The reconstruction clearly indicates that there are likely additional dead pixels in this area. \\
\begin{figure*}[!h]
\centering
  \includegraphics[width=0.19\linewidth]{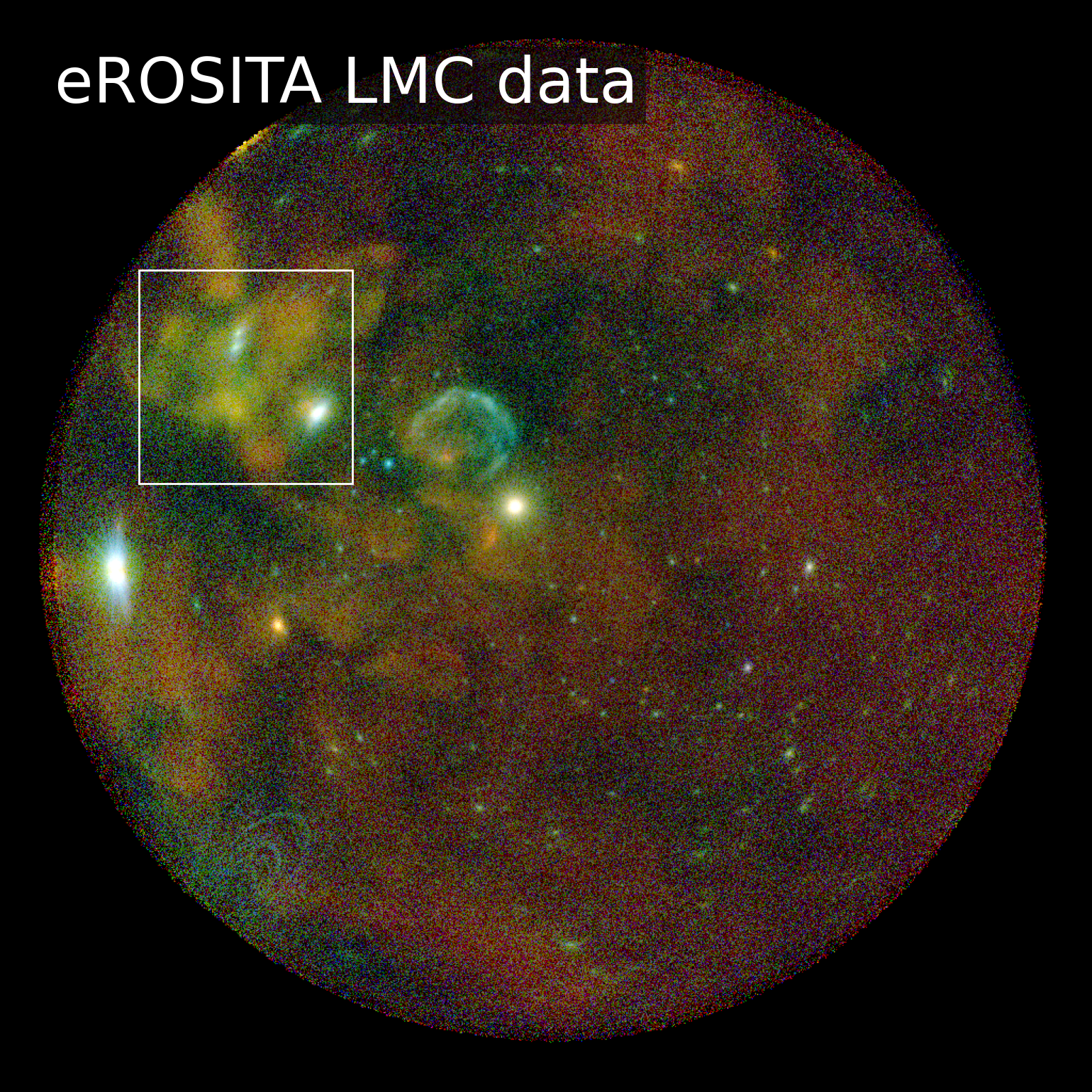}
    \includegraphics[width=0.19\linewidth]{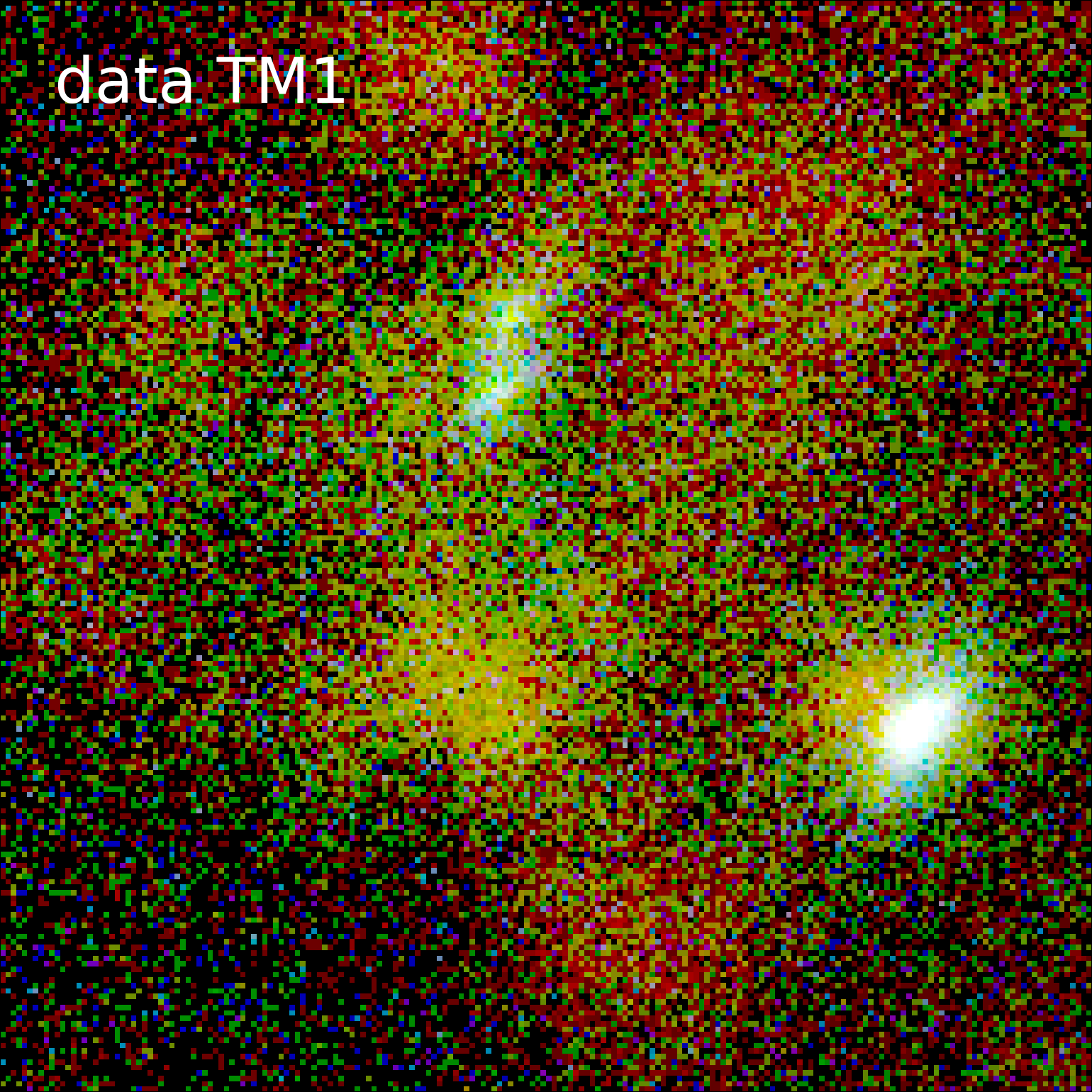}
      \includegraphics[width=0.19\linewidth]{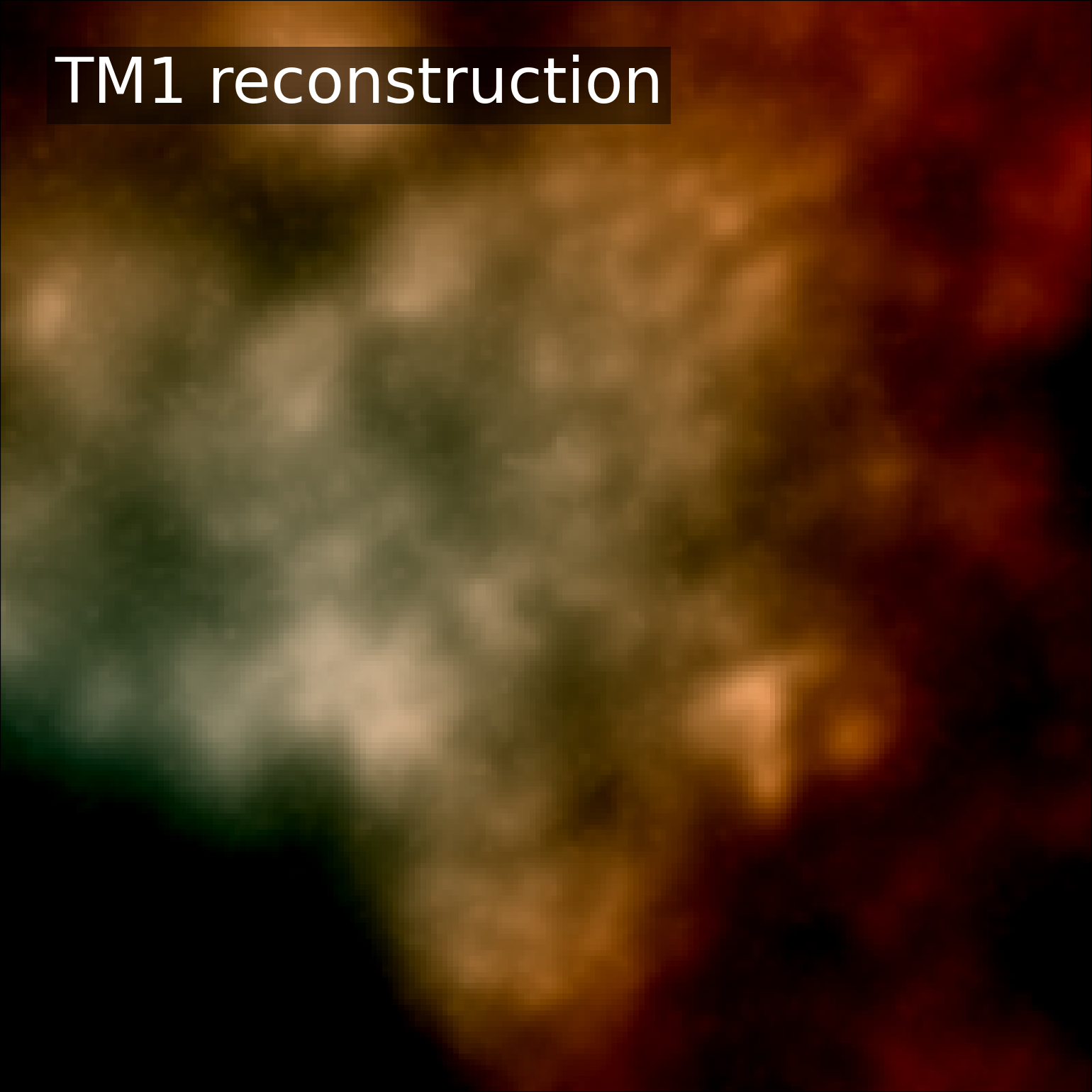}
  \includegraphics[width=0.19\linewidth]{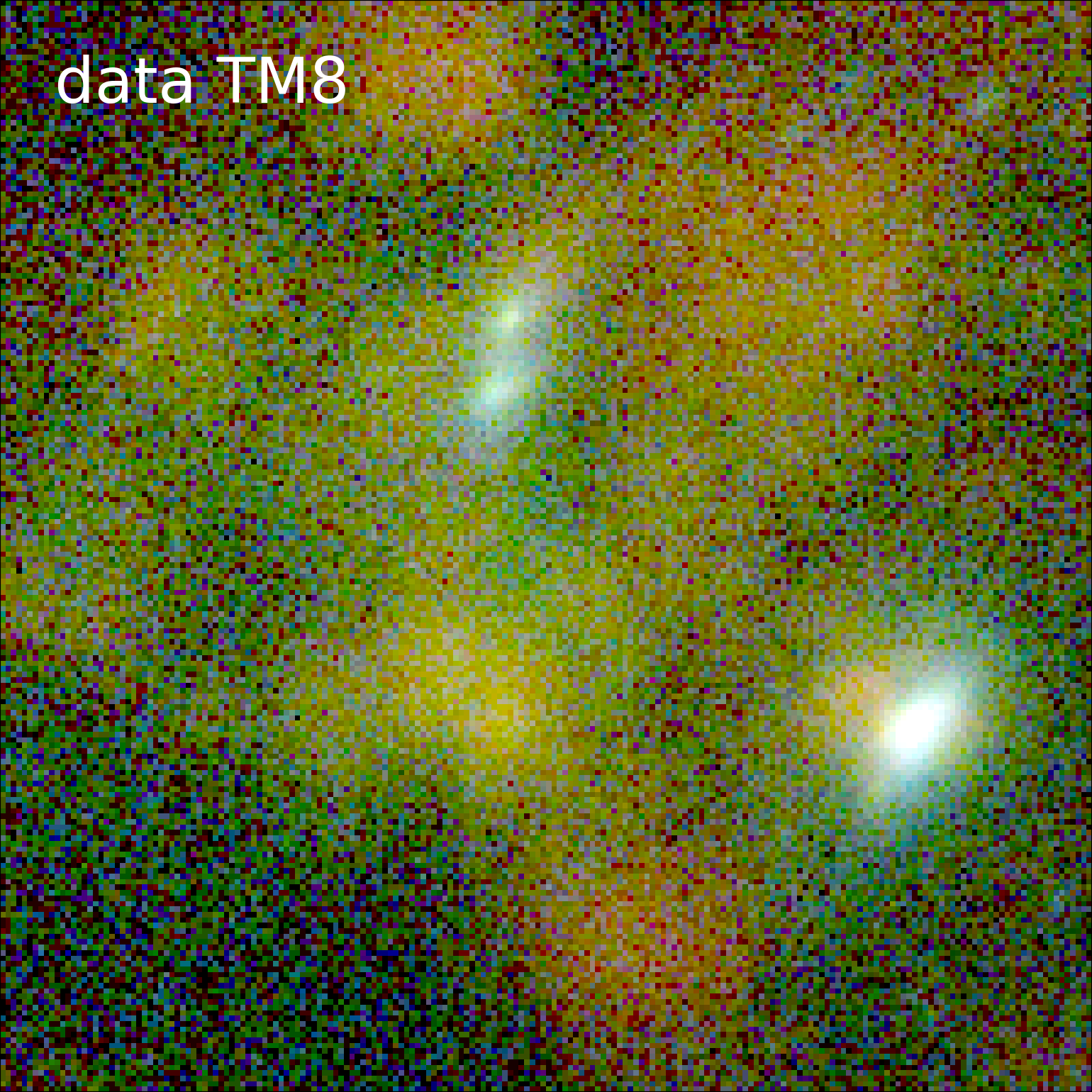}
  \includegraphics[width=0.19\linewidth]{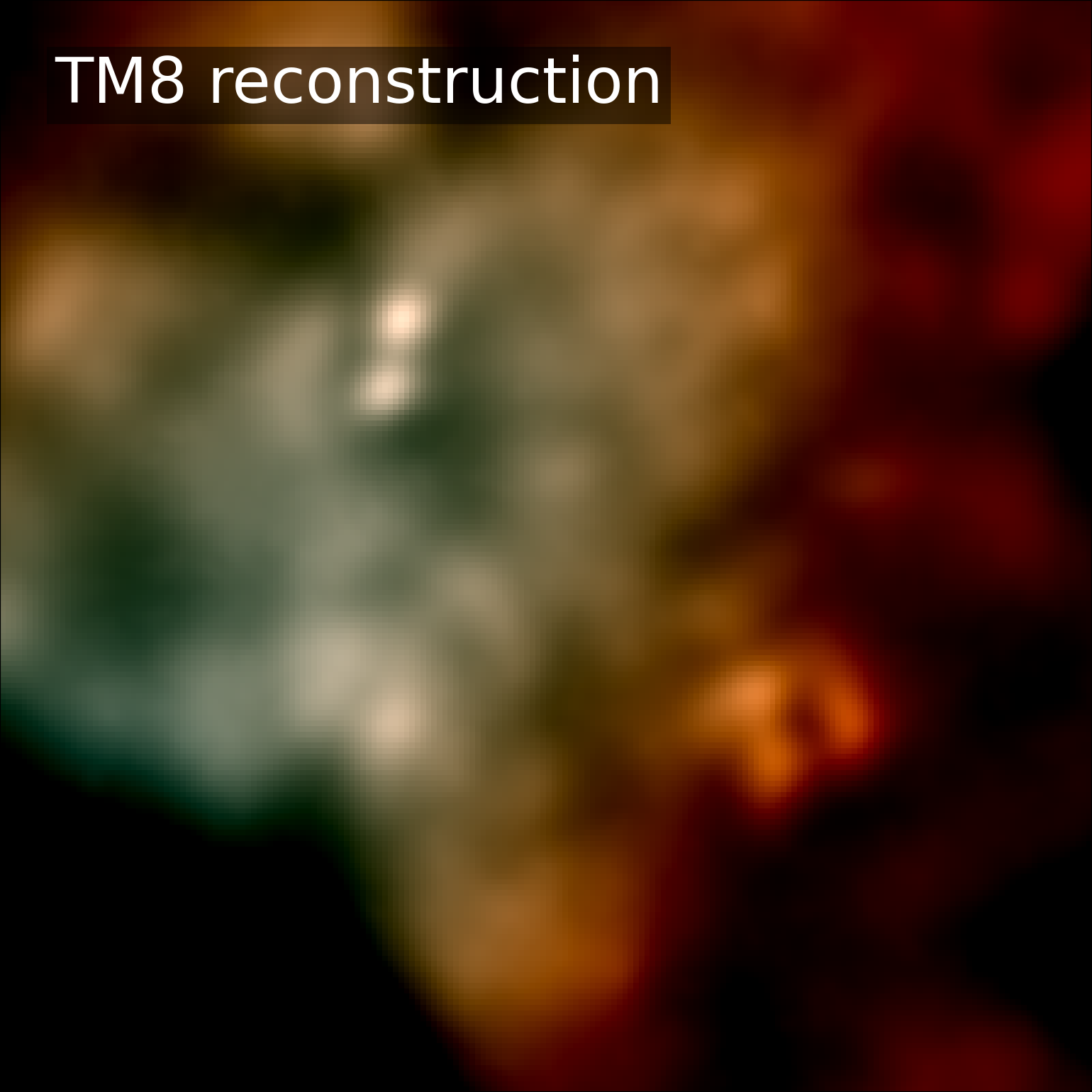}
\caption{Zoom on reconstruction of diffuse emission from the Tarantula Nebula. From left to right showing the zoom area on the plot of the eROSITA LMC data, the zoomed LMC data for \ac{TM}1, the corresponding single-\ac{TM} diffuse emission reconstruction for \ac{TM}1, the zoomed data for all 5 \acp{TM} and the diffuse emission reconstruction by means of all five observations.}
  \label{fig:zoom_images}
\end{figure*}
Figure~\ref{fig:zoom_images} shows a comparison of the diffuse structures around the Tarantula Nebula in the eROSITA data, zoomed in for both a single \ac{TM} and all five \acp{TM}.
We note that it is more difficult to separate point-like from diffuse emission using all the telescope modules. This is likely due to possible calibration inconsistencies in comparison with the single-\ac{TM} reconstructions.
We note that our derived point-source component provides a foundation for future catalog creation, although a direct comparison with the existing SRG/eROSITA catalog by \citet{merloni_catalog} is beyond the scope of this study. 
Such a comparison would require extensive calibration and validation, particularly given the calibration artifacts identified in our reconstructions and the fact that our analysis did not utilize the full available dataset across the entire field of view. 
Nevertheless, our methodology demonstrates the feasibility of automated point-source extraction from Bayesian reconstructions, highlighting a promising direction for future systematic studies and catalog validation efforts. Since in this method we define a threshold for the reconstructed point-source field, we quantify the flux discarded by the point-source thresholding process and assess its relevance in Appendix~\ref{app:resultsdiag}.

Figure~\ref{fig:ChandraData} gives a view of Chandra data for the region of interest of the \ac{LMC}, binned to $1024\times 1024$ spatial and $3$ spectral pixels. Specifically, we chose a 4 arcsec resolution to match the chosen eROSITA resolution. In Fig.~\ref{fig:ChandraData}, we plot the data using the same plotting routine and color-coding for Chandra and eROSITA. Compared to eROSITA data and reconstruction, Chandra provides finer detail due to its higher spatial resolution (\(\SI{0.5}{\arcsec}\)). This enables us to confirm that the small-scale features in this eROSITA reconstruction -- resolved through \ac{PSF} deconvolution and shot noise removal -- are real and not artifacts.

\begin{figure}[!h]
\centering
  \includegraphics[width=0.98\linewidth]{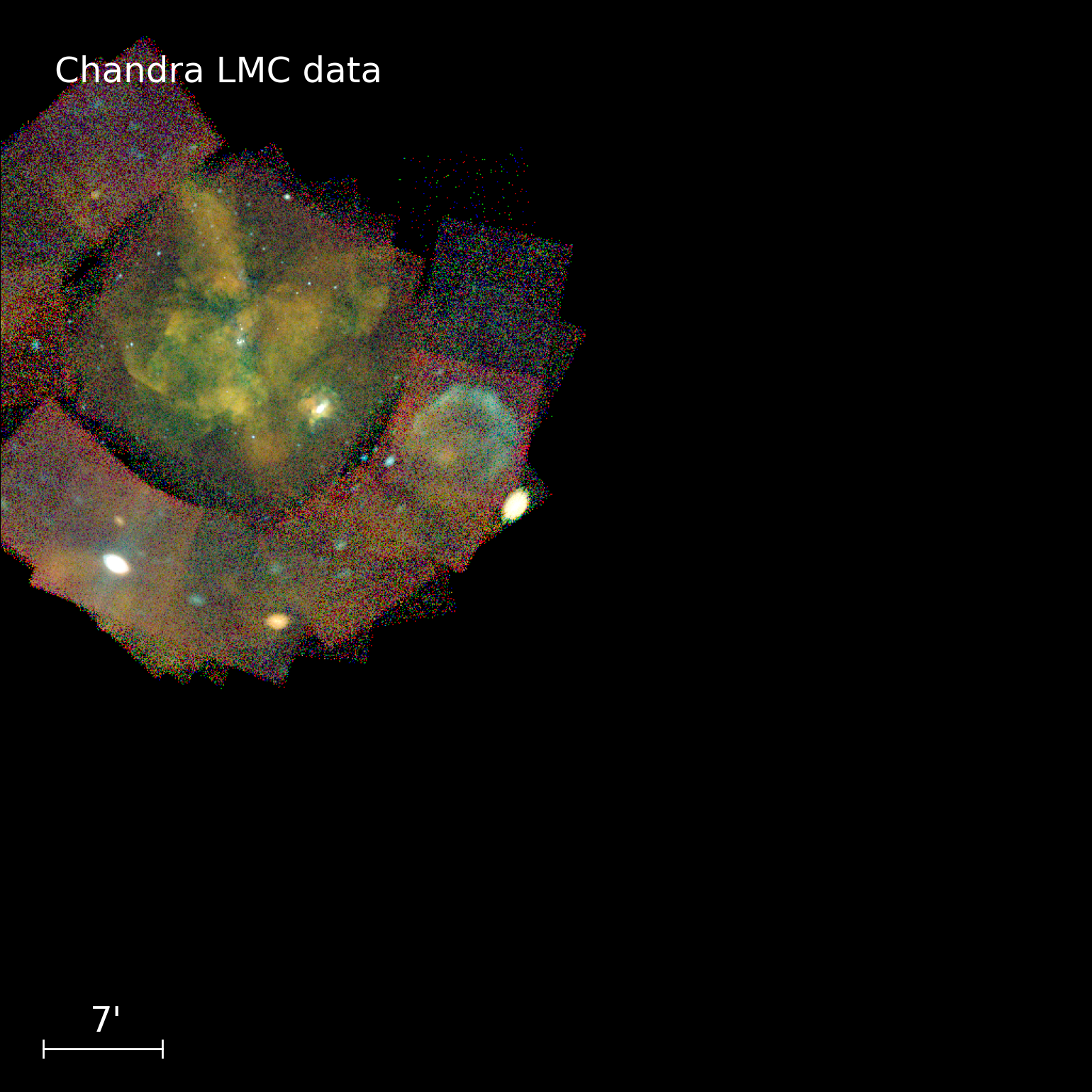}\\
  \vspace{1em}
  \includegraphics[width=0.45\linewidth]{figs/zoom_cut_rec_diffuse_rgb_tm1.png}
  \hspace{1em}
  \includegraphics[width=0.45\linewidth]{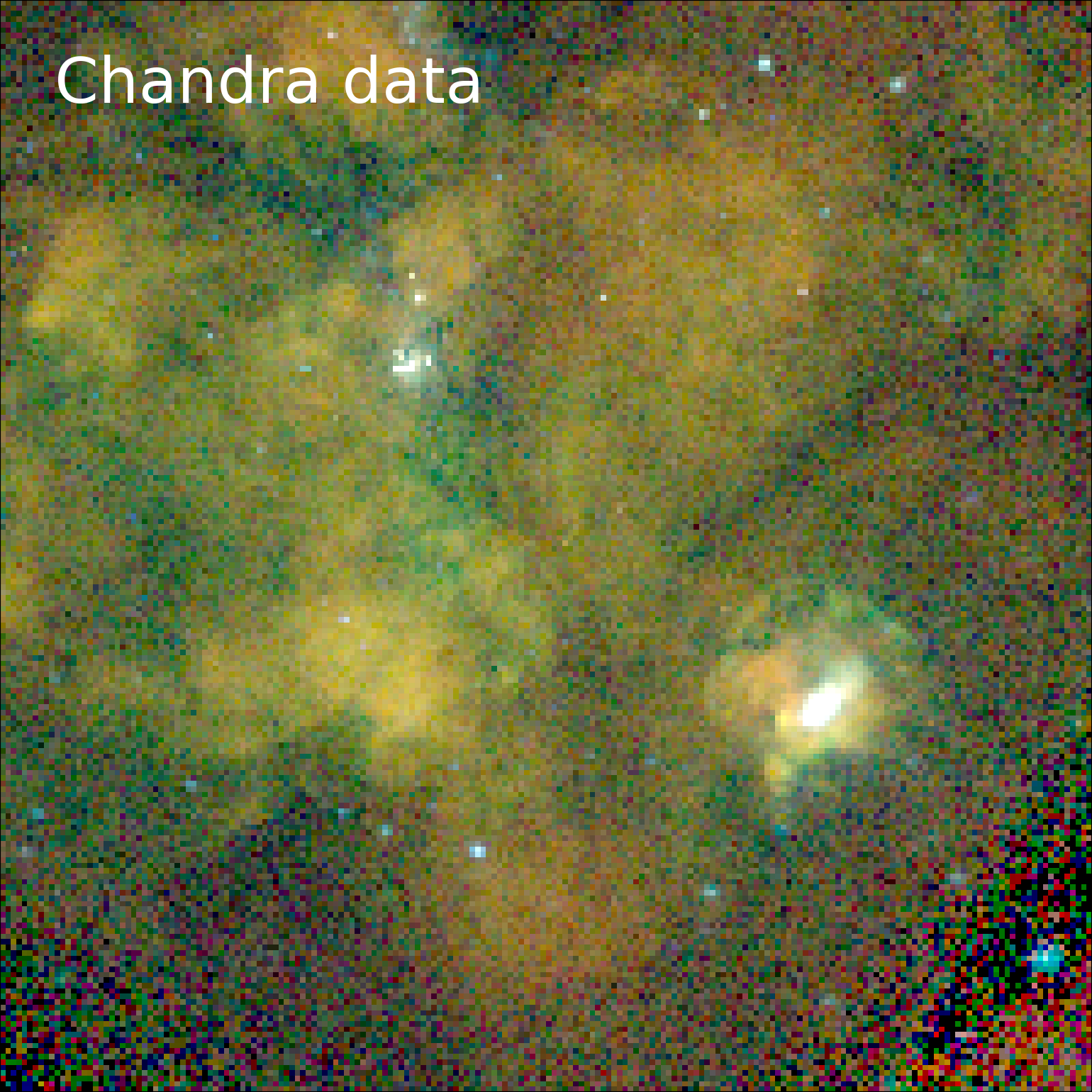}
\caption{Chandra \ac{LMC} data. \textit{The top panel} shows Chandra data, as specified in table \ref{tab:chandra_data_table}, in the region of the \ac{LMC}. \textit{The bottom panels} show the corresponding zoom-ins of our eROSITA reconstruction of the diffuse emission based on the data of \ac{TM}1 and the Chandra data on the fine-scale structures of the Tarantula Nebula, as shown in Fig.~\ref{fig:zoom_images}.}
  \label{fig:ChandraData}
\end{figure}

\section{Conclusion}
\label{sec:conclusion}
In conclusion, this paper presents the first Bayesian reconstruction of the eROSITA \ac{EDR} data, providing a denoised, deconvolved, and separated view of the diffuse and point-like sources in the \ac{LMC}.~\footnote{The reconstructed fields can be found at \url{https://doi.org/10.5281/zenodo.16918521}.} 
The presented algorithm enables the spatio-spectral reconstruction of the \ac{LMC}, incorporating its observation by the five different \acp{TM} of TM8. 
Ultimately, the reconstruction shows distinct fine-scale structures in the diffuse emission of the \ac{LMC} and deconvolved, sharp point sources, which are barely seen from the eROSITA data and verified via the comparison with higher resolved Chandra data. Thus, the presented  results have the potential to assist in the further analysis of the diffuse X-ray emission as done by \cite{Sasaki:2022} without any noise or point source contributions or effects from the \ac{PSF}. It also allows the point source catalog to be refined by considering only the point source component. Due to the generative nature of the algorithm we are able to generate simulated data, on which we tested the consistency of the reconstruction. The underlying building blocks of the implementation are publicly available \citep{JUBIK:2024} and can therefore be used to image other eROSITA observations as well. \\
The presented algorithm uses an additional component in the region of 30 Doradus C. 
Such additional components in certain regions allow to image such extended objects which overlap with the emission from the hot phase of the \ac{ISM} and point sources and have a very different correlation structure. In this way, not only the general diffuse and point source emission can be decomposed, but also the diffuse emission from the hot phase of the \ac{ISM} and from extended sources such as 30 Doradus C can be distinguished. 
In this work, the additional component for the extended source was set by hand. For future work, we aim to automate this and to find the extended sources for high excitations in the latent space. \\
There are also several areas for further investigation. The algorithm presented here can be useful to check the calibration using single \ac{TM} reconstructions and diagnostics such as the \acp{NWR}, which are readily available due to the algorithm's probabilistic nature. Future work could focus on improving the spectral resolution to allow further insight into the spectra of the different components. In addition, work is underway to extend the applicability to eROSITA field scans and all-sky surveys.
Finally, the flexibility of the algorithm extends beyond eROSITA. Its general framework can be adapted to other photon-counting observatories, such as Chandra, XMM-Newton, and more, enabling high-resolution imaging across diverse instruments. By making the instrument models publicly available through \texttt{J-UBIK}, we aim to facilitate future developments and applications, including Bayesian multi-messenger imaging of specific celestial objects.

\begin{acknowledgements}
V. Eberle, M. Guardiani, and M. Westerkamp acknowledge financial support from the German Aerospace Center and Federal Ministry of Education and Research through the project \textit{Universal Bayesian Imaging Kit -- Information Field Theory for Space Instrumentation} (F{\"o}rderkennzeichen 50OO2103).
P. Frank acknowledges funding through the German Federal Ministry of Education and Research for the project ErUM-IFT: Informationsfeldtheorie für Experimente an Großforschungsanlagen (Förderkennzeichen: 05D23EO1).

This work is based on data from eROSITA, the soft X-ray instrument aboard SRG, a joint Russian-German science mission supported by the Russian Space Agency (Roskosmos), in the interests of the Russian Academy of Sciences represented by its Space Research Institute (IKI), and the Deutsches Zentrum für Luft- und Raumfahrt (DLR). The SRG spacecraft was built by Lavochkin Association (NPOL) and its subcontractors, and is operated by NPOL with support from the Max Planck Institute for Extraterrestrial Physics (MPE). The development and construction of the eROSITA X-ray instrument was led by MPE, with contributions from the Dr. Karl Remeis Observatory Bamberg \& ECAP (FAU Erlangen-Nuernberg), the University of Hamburg Observatory, the Leibniz Institute for Astrophysics Potsdam (AIP), and the Institute for Astronomy and Astrophysics of the University of Tübingen, with the support of DLR and the Max Planck Society. The Argelander Institute for Astronomy of the University of Bonn and the Ludwig Maximilians Universität Munich also participated in the science preparation for eROSITA.

The eROSITA data shown here were processed using the eSASS software system developed by the German eROSITA consortium.

\end{acknowledgements}

\bibliographystyle{aa} % style aa.bst
\bibliography{bib.bib} % your references Yourfile.bib

\listofobjects

\begin{appendix} %First  appendix
\section{eROSITA observation of LMC SN1987A} \label{app:data}
\begin{figure}[!h]
\centering
  {TM1:}
  \includegraphics[width=\linewidth]{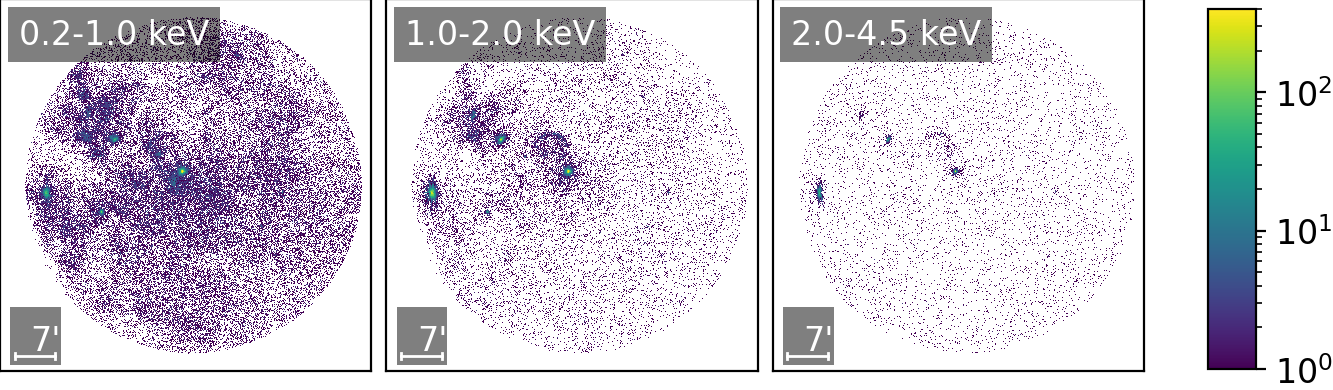} \\
  {TM2:}
  \includegraphics[width=\linewidth]{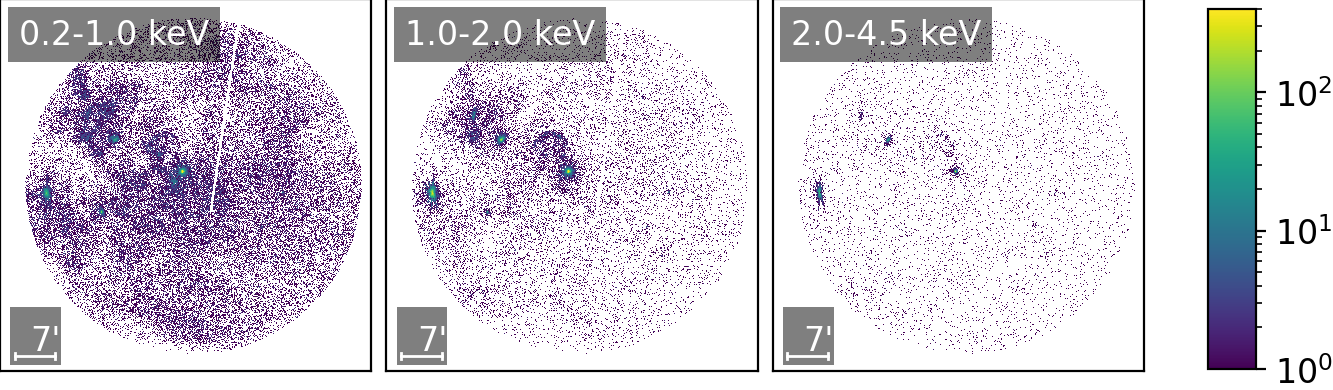} \\
{TM3:}
  \includegraphics[width=\linewidth]{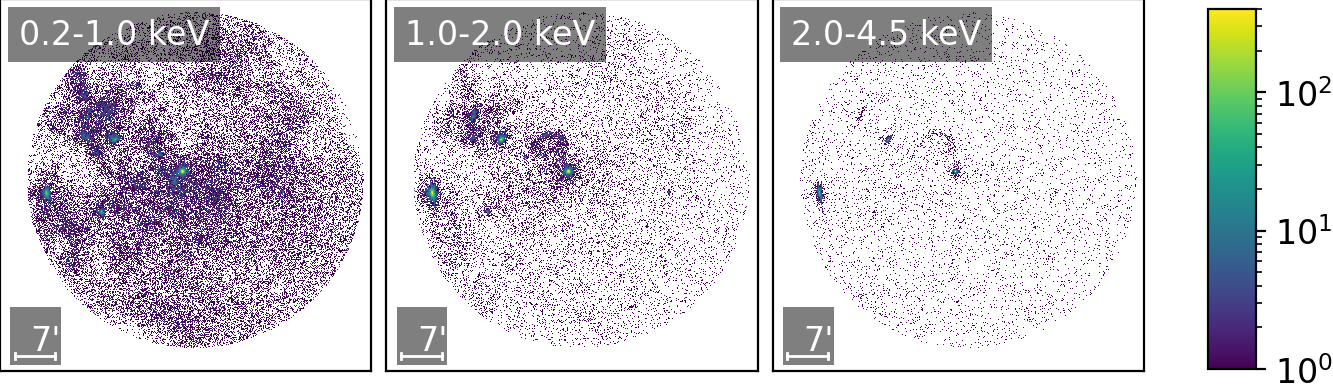} \\
{TM4:}
  \includegraphics[width=\linewidth]{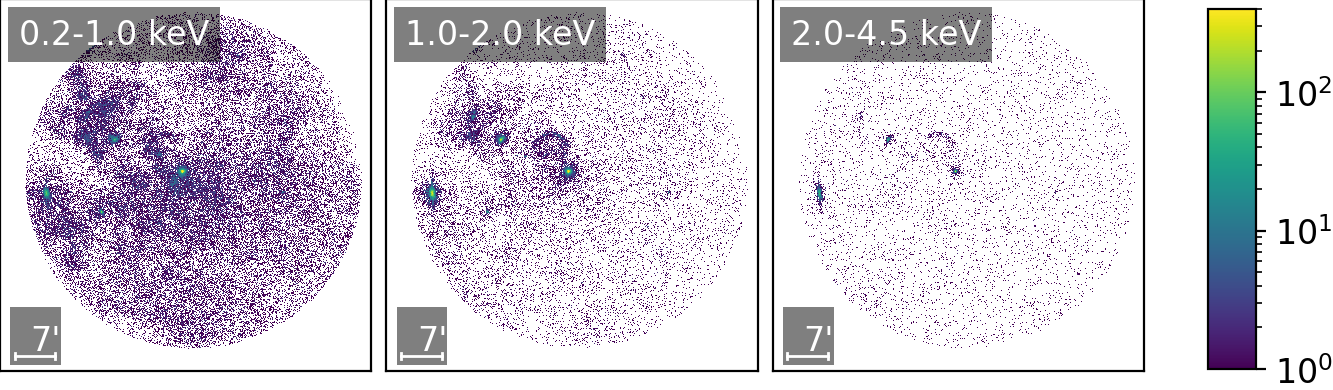} \\
{TM6:}
  \includegraphics[width=\linewidth]{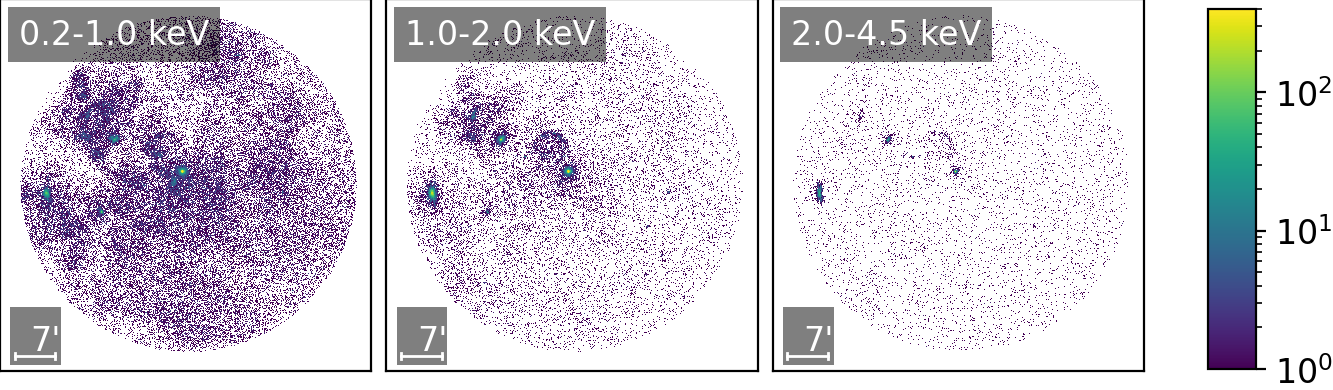} \\
\caption{Visualization of eROSITA data per energy bin in number of counts from left to right, $0.2\ \text{-}\ 0.1\; \si{\kilo\electronvolt}$, $1.0\ \text{-}\ 2.0\; \si{\kilo\electronvolt}$, and $2.0\ \text{-}\ 4.5\; \si{\kilo\electronvolt}$ for TM1 to TM6 from top to bottom.}
    \label{fig:TMeROSITAdata}
\end{figure}

\begin{figure}[!h]
  \centering
  \includegraphics[width=\linewidth]{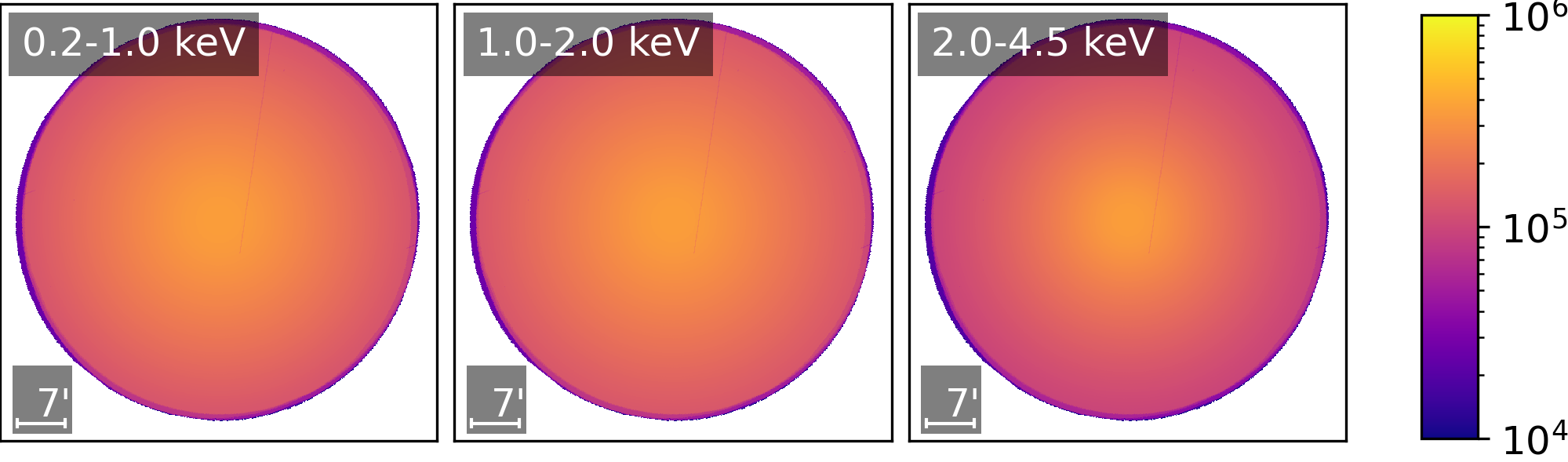} 
  \caption{eROSITA exposure maps summed over all 5 TMs.}
  \label{fig:exposures}
\end{figure}

The \ac{CalPV} data centered on SN1987A in the \ac{LMC} were pre-processed using the \ac{eSASS} pipeline, which is described in detail by \cite{Brunner2018} and \cite{Predehl2020} \footnote{Further information on the eSASS pipeline can also be found at \url{https://erosita.mpe.mpg.de/edr/DataAnalysis/}.}. In particular, the data was extracted and manipulated using the \ac{eSASS} \texttt{evtool} command. We list the flag values we chose for the \texttt{evtool} command \footnote{ A further description of the flags can be found at \url{https://erosita.mpe.mpg.de/edr/DataAnalysis/evtool_doc.html}.} in Table \ref{tab:evtoolparams}. We computed the exposure maps for the eROSITA event files using the eSASS \texttt{expmap} command and the corresponding flags in \ref{tab:expmapparams}. 
The data per energy bin and per \ac{TM} is shown in Fig.~\ref{fig:TMeROSITAdata}. 
The corresponding expsoure maps summed over the 5 \ac{TM}s are shown in Fig.~\ref{fig:exposures}
The \acp{PSF} used for the \ac{PSF} \textit{linear patched convolution} representation can be found in \texttt{tm[1-7]\_2dpsf\_190219v05.fits} 
in the \ac{CalDB}.
The effective area for the individual \ac{CA} can be found in \texttt{tm[1-7]\_arf\_filter\_000101v02.fits} in the \ac{CalDB}.
\begin{table}[!h]
    \centering
    \caption{Flags and their corresponding data types for \texttt{evtool} where tmid, where tmid $\in \{1, 2, 3, 4, 6\}$ and (emin, emax) $\in\{ (0.2, 1.0), (1.0, 2.0), (2.0, 4.5)\}$}
    \label{tab:evtoolparams}
        \centering
        \begin{tabular}{|l|l|l|}
        \hline
        \textbf{Flag} & \textbf{Data Type} & \textbf{Value}\\ \hline
        \texttt{clobber} & bool & True\\ \hline
        \texttt{events} & bool & True\\ \hline
        \texttt{image} & bool & True\\ \hline
        \texttt{size} & int & 1024\\ \hline
        \texttt{rebin} & int & 80 \\ \hline
        \texttt{center\_position} & tuple & None \\ \hline
        \texttt{region} & str & None \\ \hline
        \texttt{gti} & str & None\\ \hline
        \texttt{flag} & str & None\\ \hline
        \texttt{flag\_invert} & bool & None\\ \hline
        \texttt{pattern} & int & 15\\ \hline
        \texttt{telid} & int & tmid\\ \hline
        \texttt{emin} & float | str & emin\\ \hline
        \texttt{emax} & float | str & emax\\ \hline
        \texttt{rawxy} & str &  None\\ \hline
        \texttt{rawxy\_telid} & int & None \\ \hline
        \texttt{rawxy\_invert} & bool & False\\ \hline
        \texttt{memset} & int & None\\ \hline
        \texttt{overlap} & float & None\\ \hline
        \texttt{skyfield} & str & None\\ \hline
        \end{tabular}
\end{table}
\begin{table}[!h]
    \centering
    \caption{Flags and their corresponding data types for \texttt{expmap}, where tmid $\in \{1, 2, 3, 4, 6\}$ and (emin, emax) $\in\{ (0.2, 1.0), (1.0, 2.0), (2.0, 4.5)\}$}
        \label{tab:expmapparams}
        \centering
    \begin{tabular}{|l|l|l|}
        \hline
        \textbf{Parameter} & \textbf{Data Type} & \textbf{Value} \\ \hline
        \texttt{emin} & float | str & emin \\ \hline
        \texttt{emax} & float | str & emax\\ \hline
        \texttt{withsinglemaps} & bool & True\\ \hline
        \texttt{withmergedmaps} & bool & False\\ \hline
        \texttt{gtitype} & str & GTI\\ \hline
        \texttt{withvignetting} & bool & True \\ \hline
        \texttt{withdetmaps} & bool & True \\ \hline
        \texttt{withweights} & bool & True\\ \hline
        \texttt{withfilebadpix} & bool & True \\ \hline
        \texttt{withcalbadpix} & bool & True\\ \hline
        \texttt{withinputmaps} & bool & False \\ \hline
    \end{tabular}
\end{table}

\section{Chandra observations of the LMC} \label{app:chandra_data}
In Fig.~\ref{fig:ChandraData} the Chandra data of certain regions of the \ac{LMC} is shown. The according observations, which were taken into account, are specified in this section in table \ref{tab:chandra_data_table}.
\begin{table}
\centering
\begin{tabular}{ccccc} \toprule
\bfseries ObsID & \bfseries Inst. & \bfseries R.A. & \bfseries Decl. & \bfseries Date\\ \midrule
22 & ACIS-I & 05:38:42.9  & -69:06:03.0 & 21.9.1999 \\
5906 & ACIS-I & 05:38:42.4 & -69:06:02.0 & 21.1.2006 \\
7263 & ACIS-I & 05:38:42.4 & -69:06:02.0 & 22.1.2006 \\
7264 & ACIS-I & 05:38:42.4 & -69:06:02.0 & 30.1.2006 \\
16192 & ACIS-I & 05:38:42.4 & -69:06:02.9 & 03.5.2014 \\
16193 & ACIS-I & 05:38:42.4 & -69:06:02.9 & 08.5.2014 \\ 
16194 & ACIS-I & 05:38:42.4 & -69:06:02.9 & 12.5.2014 \\
16195 & ACIS-I & 05:38:42.4 & -69:06:02.9 & 24.5.2014 \\
16196 & ACIS-I & 05:38:42.4 & -69:06:02.9 & 30.5.2014 \\
16197 & ACIS-I & 05:38:42.4 & -69:06:02.9 & 06.6.2014 \\
16198 & ACIS-I & 05:38:42.4 & -69:06:02.9 & 11.6.2014 \\
16199 & ACIS-I & 05:38:42.4 & -69:06:02.9 & 27.3.2015 \\
16200 & ACIS-I & 05:38:42.4 & -69:06:02.9 & 26.6.2014 \\
16201 & ACIS-I & 05:38:42.4 & -69:06:02.9 & 21.7.2014 \\
16202 & ACIS-I & 05:38:42.4 & -69:06:02.9 & 19.8.2014 \\
16203 & ACIS-I & 05:38:42.4 & -69:06:02.9 & 02.9.2014 \\
18670 & ACIS-I & 05:38:42.4 & -69:06:02.9 & 21.1.2016 \\\bottomrule
\end{tabular}
\caption{Information on the Chandra ACIS observations of the 30 Doradus region in the \ac{LMC} used for the comparison in Fig. \ref{fig:ChandraData}}
\label{tab:chandra_data_table}
\end{table}

% \clearpage
\section{Hyperparameter Search}
\label{app:hyperparams}
The prior model described in Sect.~\ref{subsec:prior_models} requires choosing a set of hyperparameters, which describe the mean and standard deviation of the Gaussian processes modelling the prior. 
The meaning of the specific hyperparameters of the correlated field is more precisely described in \cite{Arras_2022}. 
In particular, the offset mean of the correlated field parametrizes the mean of $\tau$ in Eq~\ref{eq:gauss_process} and therefore the mean of the log-normal flux. 
Accordingly, we take the exposure-corrected data, $d_e$, shown in Fig.~\ref{fig:Data} and calculate its mean $\expval{d_e}$ and set both the offset mean of $\varphi_{\ln}$ and $\varphi_{\ln, b}$ to $\log{\expval{d_e}} = -19.9$. \\
We use the information on the detection threshold to set the hyperparameters for the inverse gamma distribution used for the point sources. 
In particular we set the mean $m$ of the inverse gamma distribution as the sum of all fluxes from point sources which are higher than the detection threshold $\theta$ divided by the total number of pixels. \\
To determine the detection threshold $\theta$, we set a minimum \ac{S/N}, $\gamma_\text{min}$, that is required to reliably detect a source. Essentially, for a source to be detected, the \ac{S/N} $\gamma$ in each pixel must be higher than this set threshold, $\gamma_\text{min}$. Specifically, for Poisson data, the \ac{S/N} is given by $\gamma = \sqrt{\lambda}$, where $\lambda$ is the expected number of counts in a pixel.
We set $\gamma_\text{min}$ based on the confidence level we want for detection. In this case, we aim for a $99\%$ confidence level, meaning there is a $99\%$ probability that any observed signal is not just a random fluctuation,
\begin{align}
    \mathcal{P}(k \geq 1|\lambda) = 1 - \mathcal{P}(k = 0|\lambda) = 1 - e^{-\lambda} \overset{!}{=} 0.99,
\end{align}
which leads to $\lambda_\text{min} = 4.6$ and, consequently, $\gamma_\text{min} = \sqrt{\lambda_\text{min}} = 2.14$.
The pixel-wise detection threshold $\theta_i$ is then defined, via the smallest flux, which can be reliably detected in each pixel $i$, which is given via $\lambda_\text{min}$ and the exposure in the corresponding pixel, $E_i$
\begin{align}
    \theta_i = \frac{\lambda_\text{min}}{E_i}.
\end{align}
These $\theta_i$'s are used as a pixel wise criterion for the acceptance of a point source in the final plots.
This line of thought can also be used in order to set priors for the inverse gamma component. Therefore, we want to find an overall detection threshold for the whole image, which is then defined via maximal exposure, $E_\text{max}$
\begin{align}
    \theta = \frac{\lambda_\text{min}}{E_\text{max}} = 2.5 \times 10^{-9}.
    \label{eq:detection_threshold}
\end{align}
Eventually, this leads to a mean $m=2.08 \times 10^{-9}$ of the inverse gamma distribution. 
The mode $M$ of the inverse gamma distribution should be even further below the detection threshold. 
In particular, we thus assume that the \ac{S/N} ratio for the mode is much lower, i.e. $\gamma_\text{min} = 0.1$,
\begin{align}
    M = \frac{0.1^2}{E_\text{max}}.
\end{align}
Having the mean and the mode, we can use these in order to calculate the hyper-parameters $\alpha$ and $q$ of the inverse gamma distribution via 
\begin{align}
    \alpha &= \frac{2}{\frac{\mu}{M}-1}+1,\\
    q      &= M (\alpha +1).
\end{align}
A prior sample drawn from the prior given these hyperparameters can be seen in Fig.~\ref{fig:Priorsample}.

\clearpage
\section{Results diagnostics}
\label{app:resultsdiag}
\begin{figure}[!h]
  \centering
    {TM1:}\\
  \includegraphics[width=0.85\linewidth]{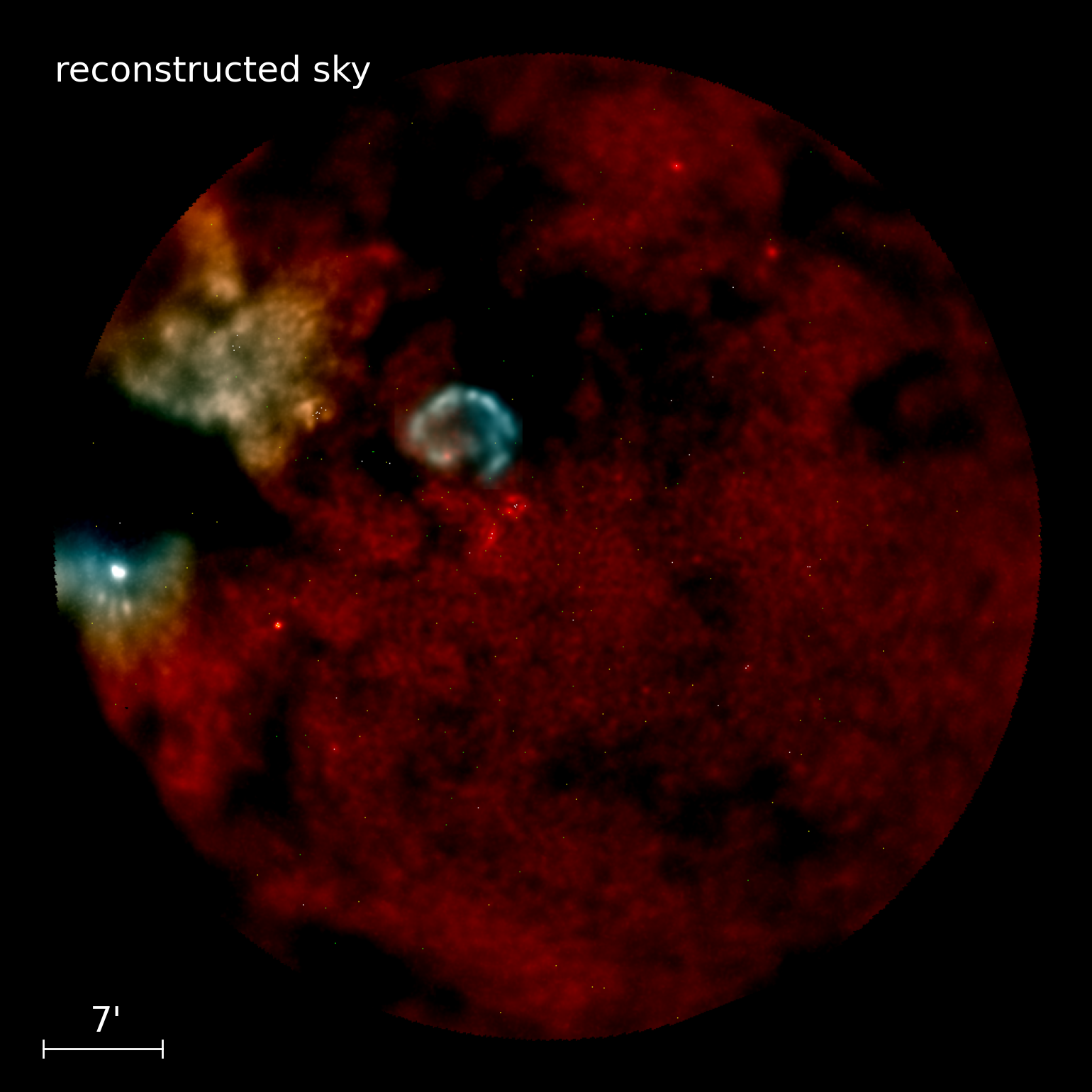}\\
    {TM2:}\\
  \includegraphics[width=0.85\linewidth]{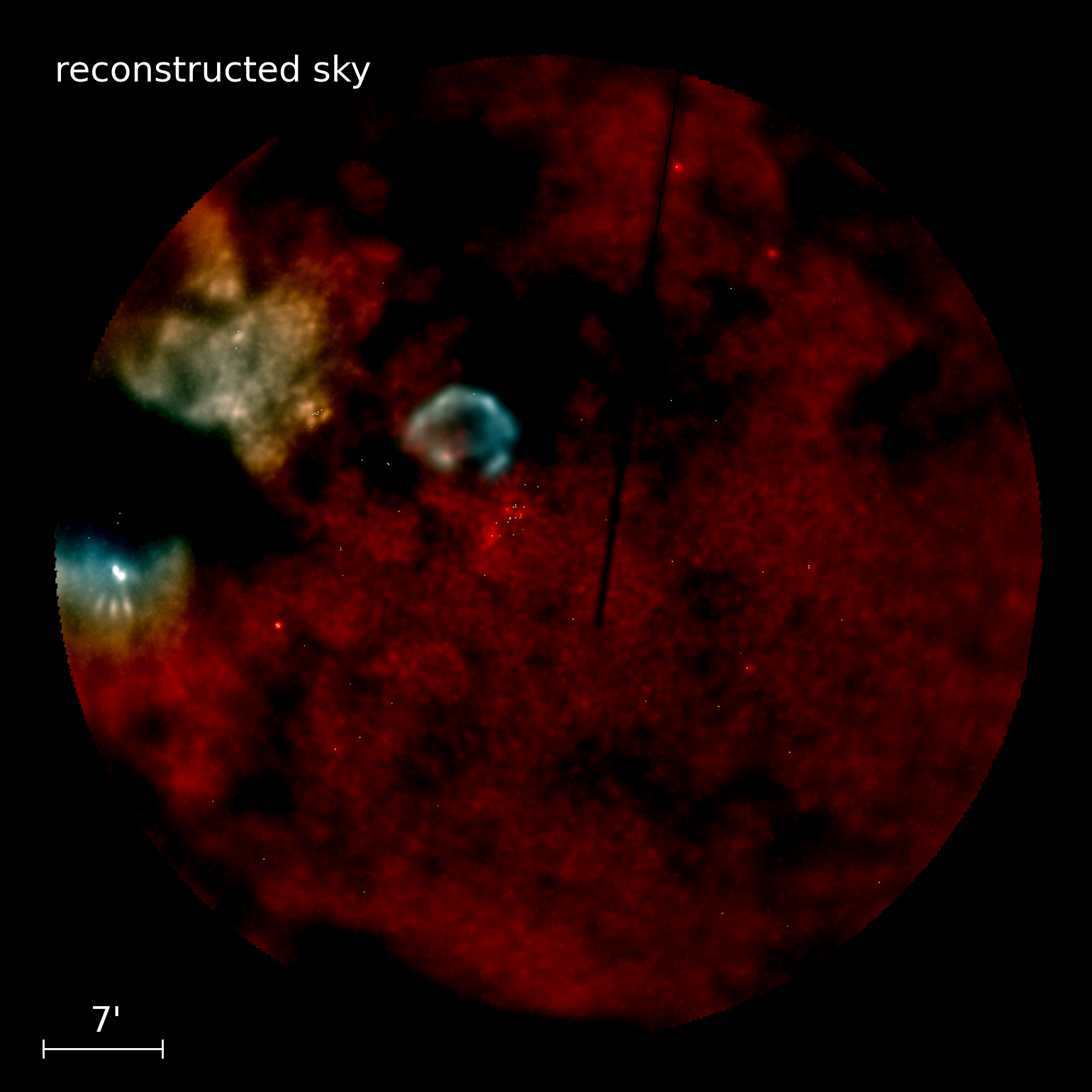}
  \caption{Results for single-TM reconstructions for \ac{TM}1, \ac{TM}2, \ac{TM}3, \ac{TM}4, \ac{TM}6.  The different colors represent the  logarithmic intensities in the three energy channels 0.2 - 0.1 keV, 1.0 - 2.0 keV,
    and 2.0 - 4.5 keV and are depicted in red, green, and blue, respectively.}
      \label{fig:rec_tm1}
\end{figure}

\begin{figure*}[!h]
  \centering
  \includegraphics[width=\linewidth]{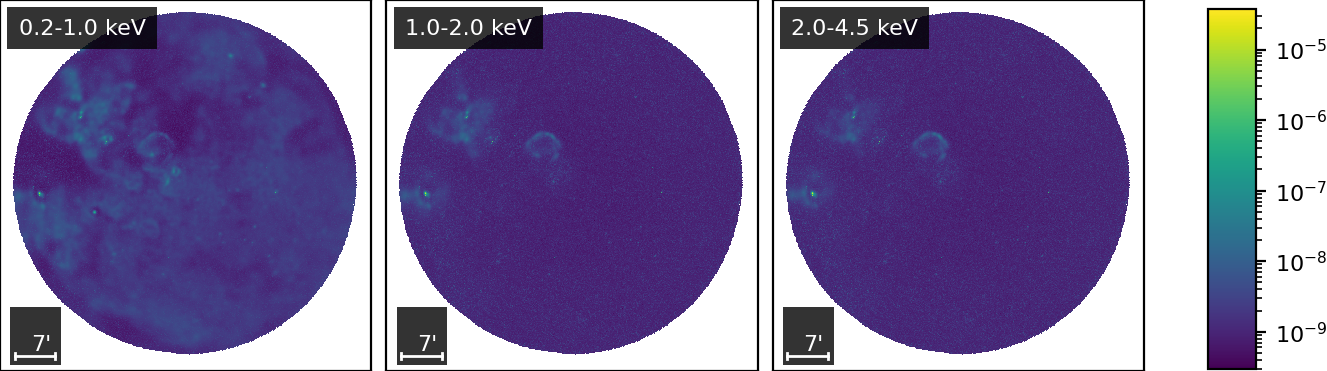}
    \caption{Posterior mean of the sky reconstruction in the different energy bins in $[1/(\arcsec^2 \times \  \mathrm{s})]$.}
 \label{fig:rec_per_energybin}
\end{figure*}

\begin{figure*}[!h]
  \centering
    {TM3:}\\
  \includegraphics[width=\linewidth]{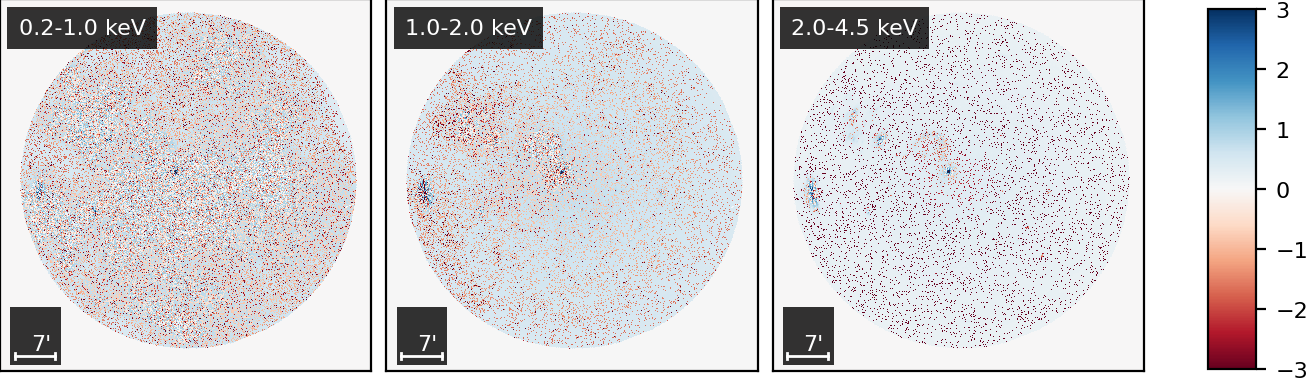}\\
    {TM4:}\\
\includegraphics[width=\linewidth]{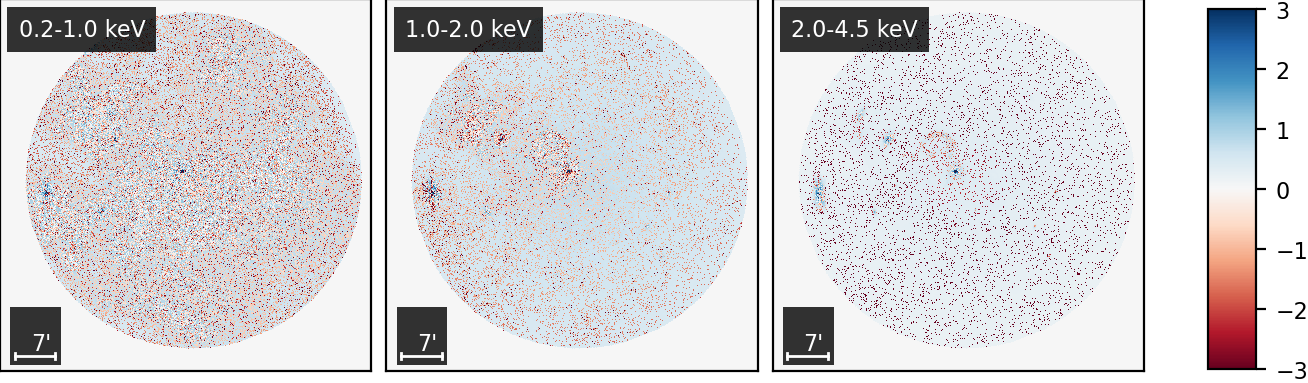}\\
    {TM6:}\\
\includegraphics[width=\linewidth]{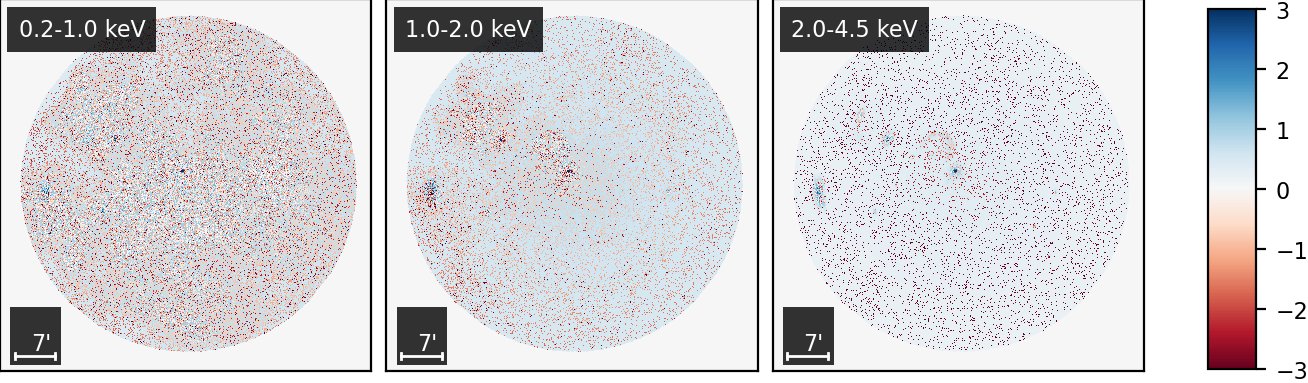}
    \caption{Posterior mean of the \acp{NWR} for single-\ac{TM} reconstructions. A value of 1 indicates that the observed data counts lie within one Poisson standard deviation of the reconstructed expected flux $\lambda$.}
    \label{fig:nwr_tm_2}
\end{figure*}

\begin{figure*}[!h]
\centering
\includegraphics[width=0.94\linewidth]{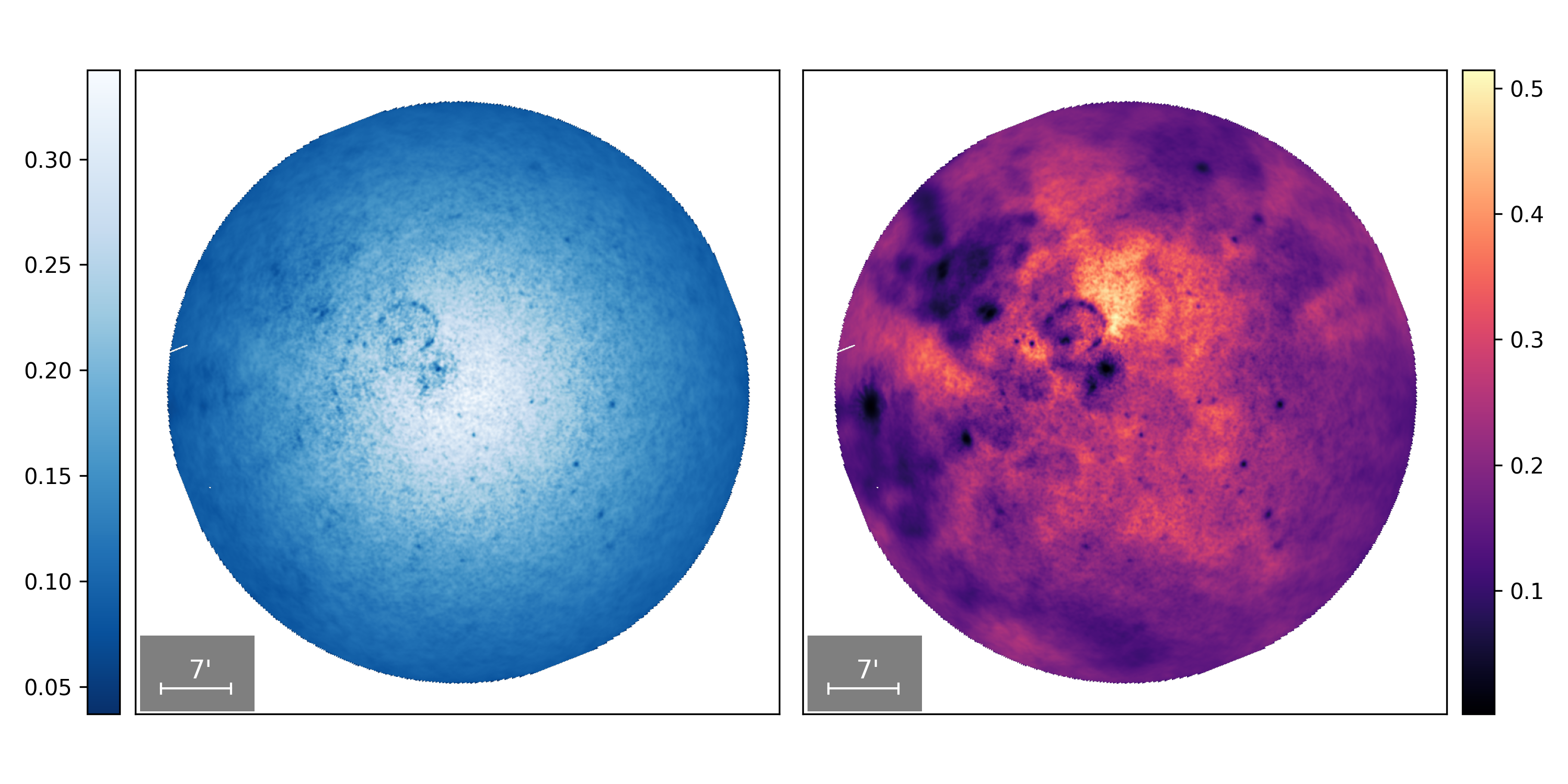}
\caption{Thresholded point-source expected count rate analysis for TM1. \textit{Left panel:} The expected count-rate ($\lambda_t$) from point-source emission below the detection threshold (Eq.~\eqref{eq:detection_threshold}), illustrating that it is everywhere below unity, indicating negligible discarded flux. \textit{Right panel:} The ratio between the thresholded point-source expected count rate ($\lambda_t$) and the expected noise ($\sqrt{\lambda}$). The flux discarded by thresholding is consistently below half of the noise level, confirming its low significance.}
\label{fig:ps_cut}
\end{figure*}

Here, we show additional plots corresponding to the reconstruction of SN1987A in the \ac{LMC} as seen by SRG/eROSITA in the \ac{CalPV} phase. 
First, we display the reconstruction shown in Fig.~\ref{fig:full_rec} per energy bin in Fig.~\ref {fig:rec_per_energybin} to give a better understanding of the color bar used. We also show the corresponding uncertainty in the form of the standard deviation
per energy bin in Fig.~\ref{fig:rec_unc_per_energybin}. 
Finally, we also performed single \ac{TM} reconstructions, the results of which are shown per \ac{TM} in Fig. \ref{fig:rec_tm1}. Important diagnostic measures to check for possible calibration inconsistencies in the single \ac{TM} reconstructions are the \acp{NWR} (Eq.~(\ref{eq:nwr})), which are shown per \ac{TM} and energy bin in Fig.~\ref{fig:nwr_tm_1} and Fig.~\ref{fig:nwr_tm_2}.
Finally, in Fig.~\ref{fig:ps_cut}, we provide a quantitative assessment of the flux discarded when applying the point-source detection threshold defined in Eq.~\eqref{eq:detection_threshold}, focusing on TM1. The left panel shows the expected count rate $\lambda_t$ associated with point-source emission below the detection threshold. We find that $\lambda_t$ remains below unity across the field, indicating that no substantial flux is removed. This is further supported by the right panel, which shows the ratio of $\lambda_t$ to the expected noise level, $\sqrt{\lambda}$, at each pixel. The thresholded flux never exceeds half the local noise level, confirming that the discarded flux is negligible and does not impact the overall reconstruction quality.

\clearpage

\begin{figure}[!h]
  \centering
         {TM3:}\\
  \includegraphics[width=0.85\linewidth]{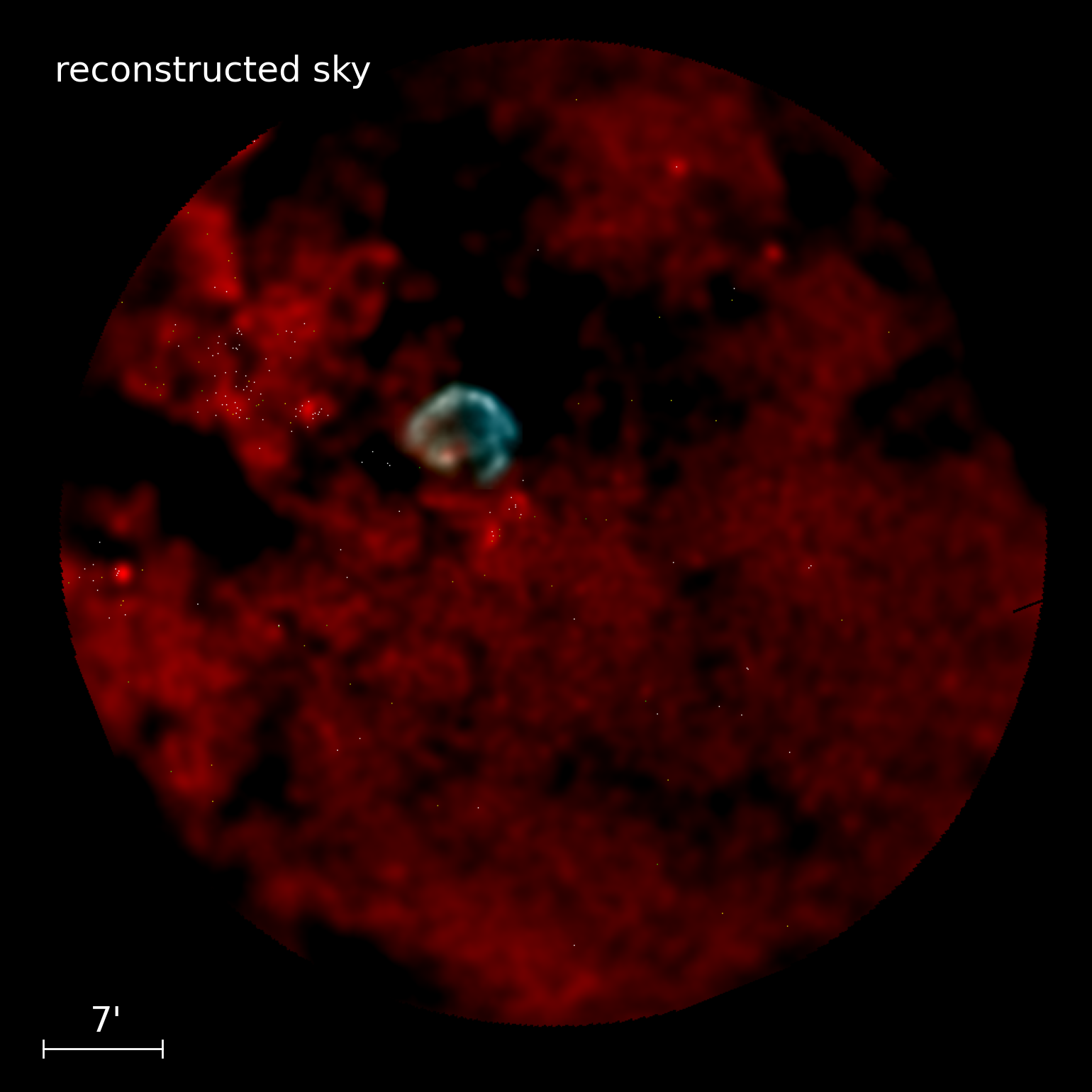}\\
    {TM4:}\\
  \includegraphics[width=0.85\linewidth]{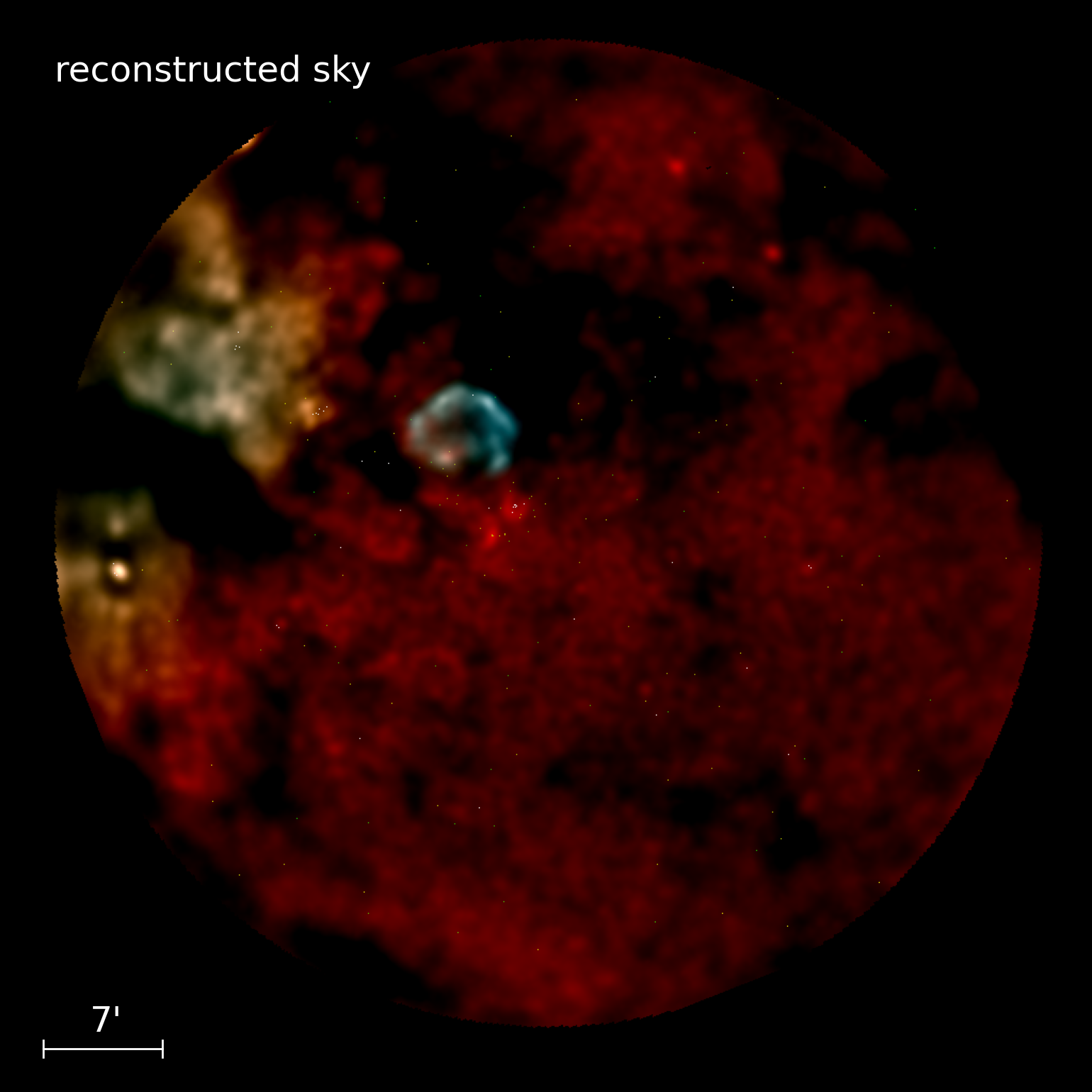} \\
    {TM6:}\\
  \includegraphics[width=0.85\linewidth]{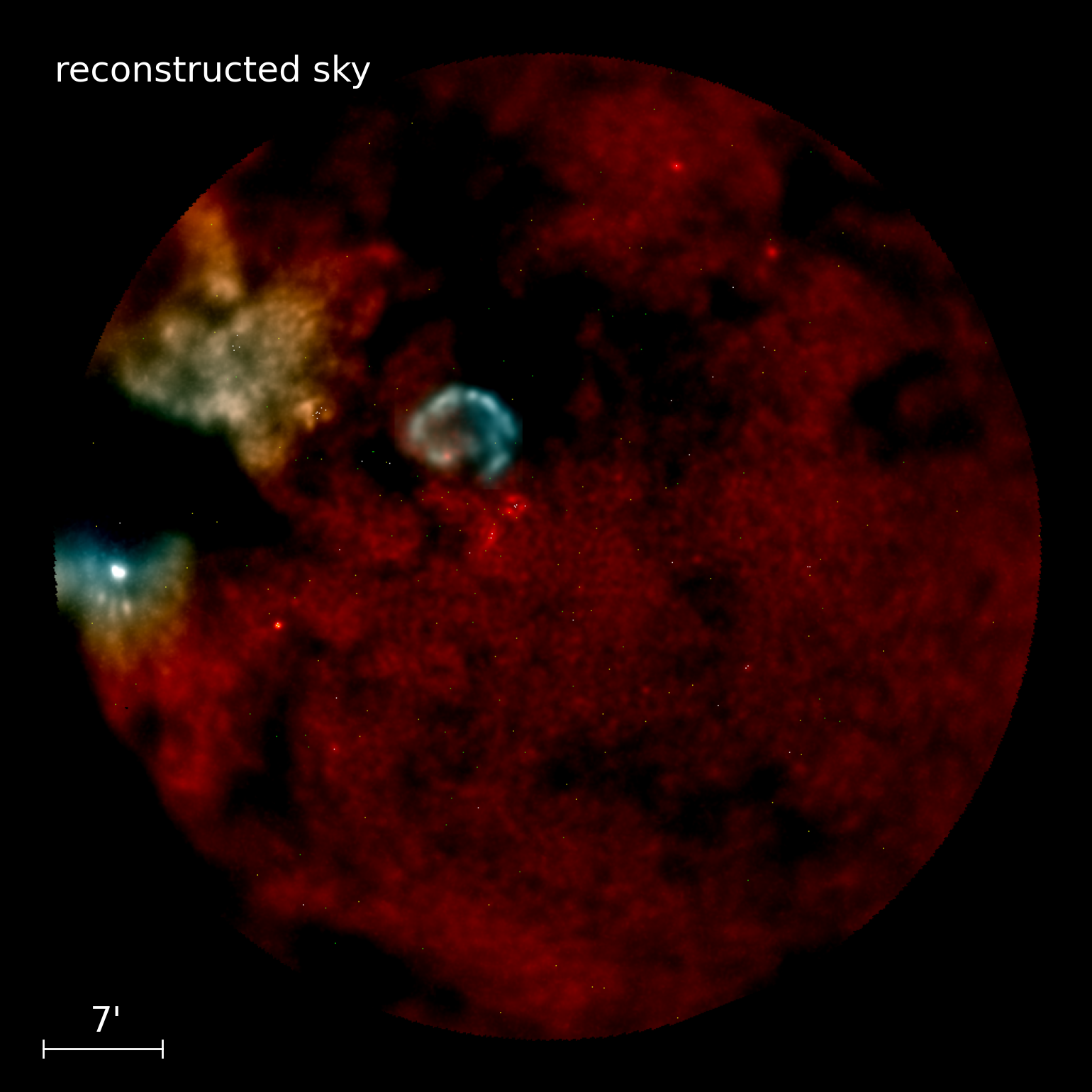}
\end{figure}

\begin{figure*}[!h]
  \centering
  \includegraphics[width=\linewidth]{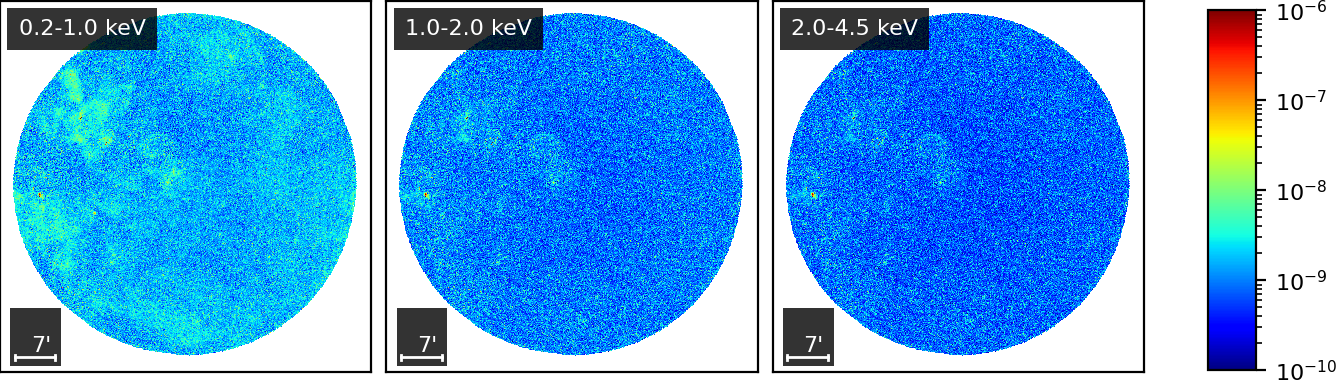}
    \caption{Standard deviation per energy bin for the reconstruction shown in Figs. \ref{fig:full_rec}+\ref{fig:rec_per_energybin} in $[1/(\arcsec^2 \times \  \mathrm{s})]$.}
    \label{fig:rec_unc_per_energybin}
\end{figure*}

\begin{figure*}[!h]
  \centering
    {TM1:}\\
  \includegraphics[width=\linewidth]{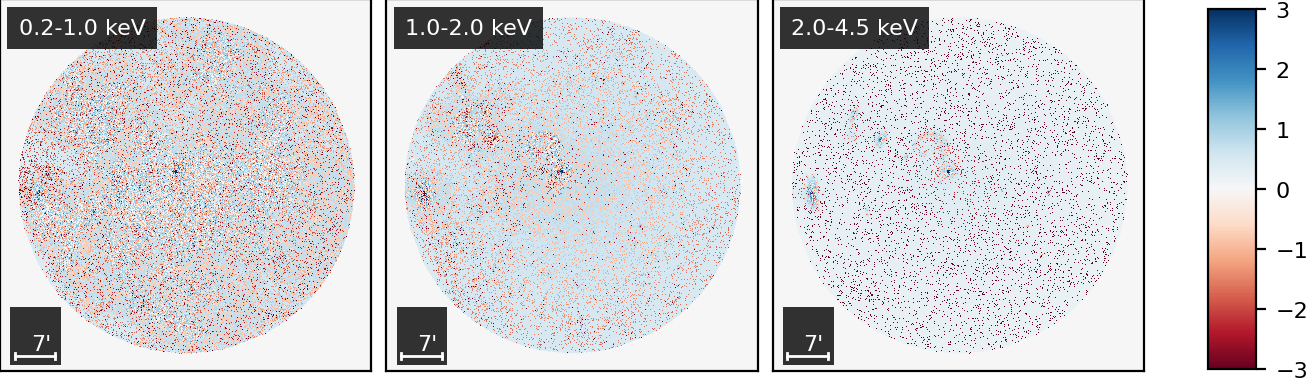}\\
    {TM2:}\\
  \includegraphics[width=\linewidth]{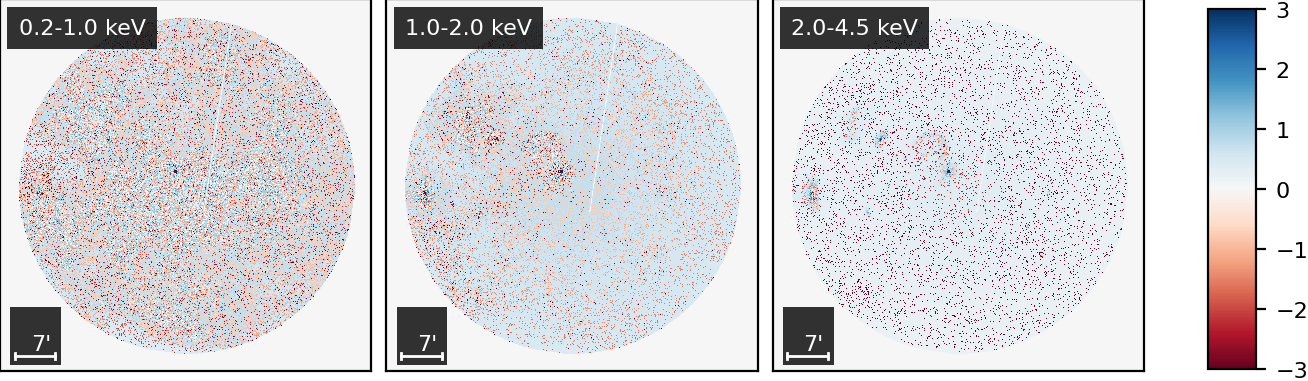}
  \caption{Continued in Fig.~\ref{fig:nwr_tm_2}.}
      \label{fig:nwr_tm_1}
\end{figure*}

\clearpage
\section{Validation diagnostics}
\label{app:validationdiagnostics}
\begin{figure}[!h]
\centering
{TM1:}
  \includegraphics[width=\linewidth]{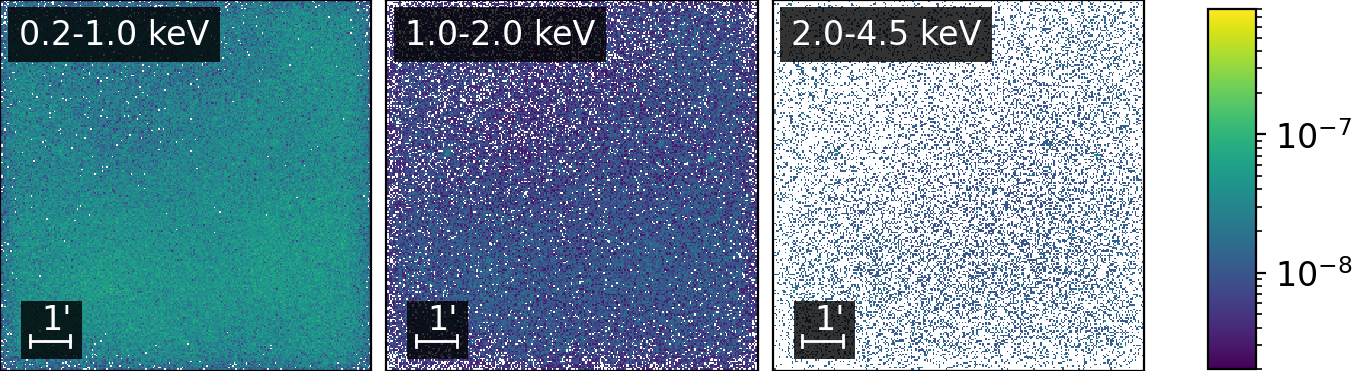} \\
{TM2:}
  \includegraphics[width=\linewidth]{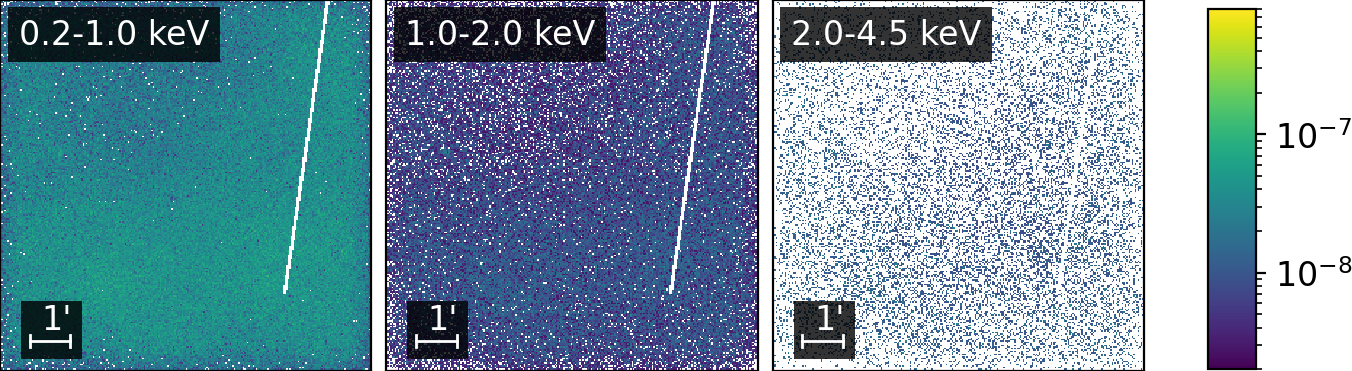} \\
{TM3:}
  \includegraphics[width=\linewidth]{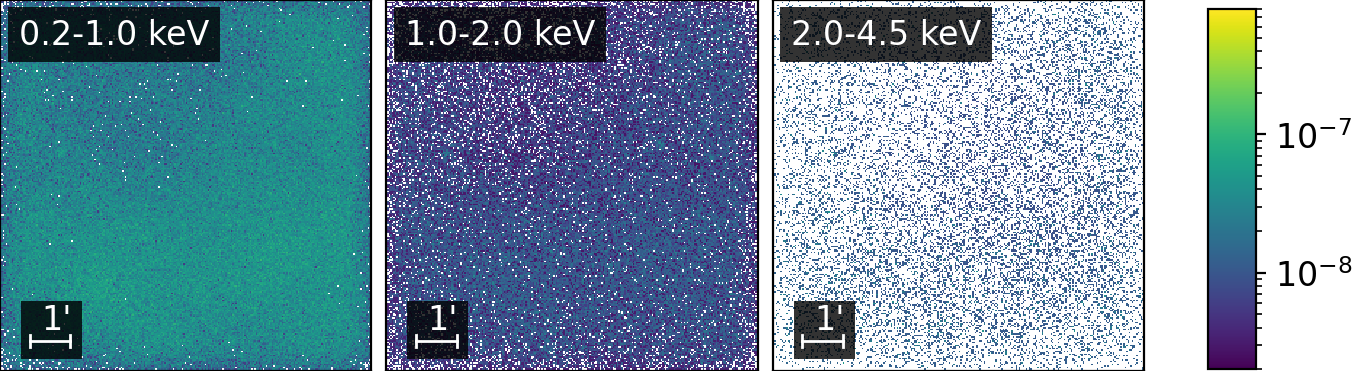} \\
{TM4:}
  \includegraphics[width=\linewidth]{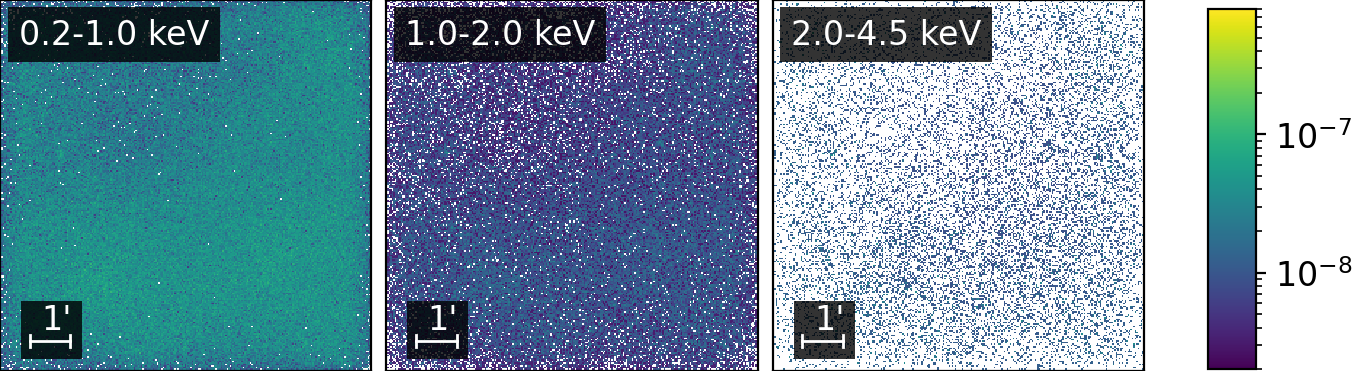} \\
{TM6:}
  \includegraphics[width=\linewidth]{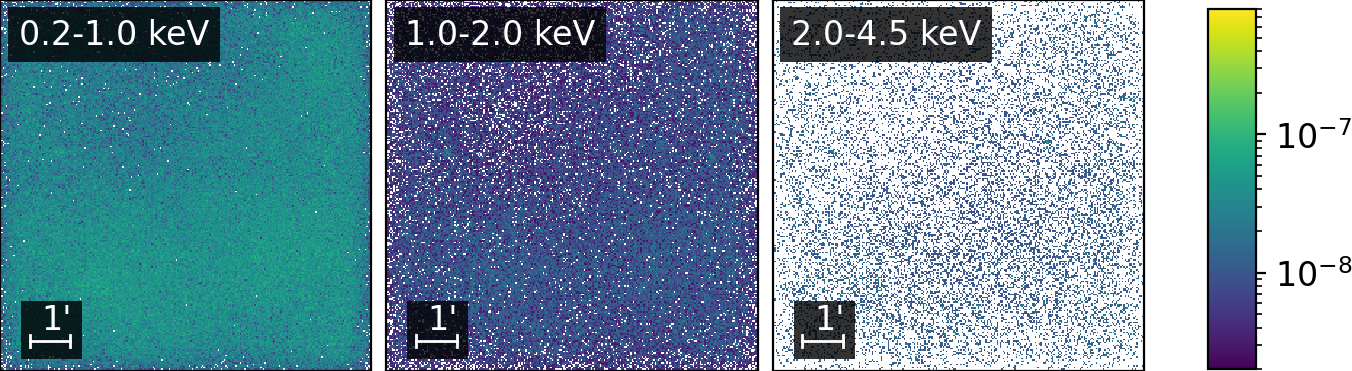} \\
\caption{Visualization of the exposure corrected simulated data per energy bin from left to right, 0.2-1.0 keV, 1.0-2.0 keV and 2.0-4.5 keV for TM1 to TM6 from top to bottom.}
  \label{fig:simdata_ebin}
\end{figure}

\begin{figure*}[!h]
\centering
  \includegraphics[width=0.94\linewidth]{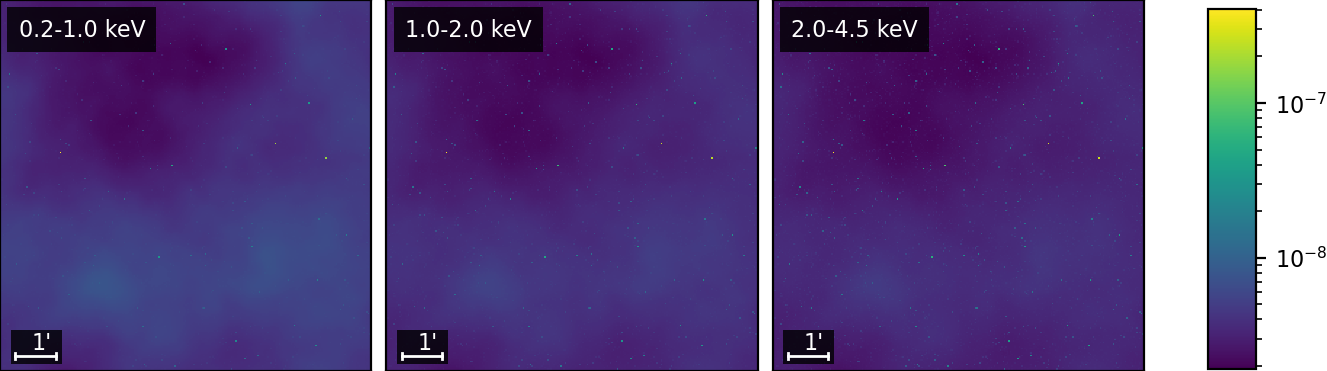} \\
\caption{Visualization of the simulated sky per energy bin (left: 0.2-1.0 keV, center: 1.0-2.0 keV, right: 2.0-4.5 keV) in $[1/(\arcsec^2 \times \  \mathrm{s})]$.}
  \label{fig:simsky_ebin}
\end{figure*}

\begin{figure*}[!h]
\centering
  \includegraphics[width=0.94\linewidth]{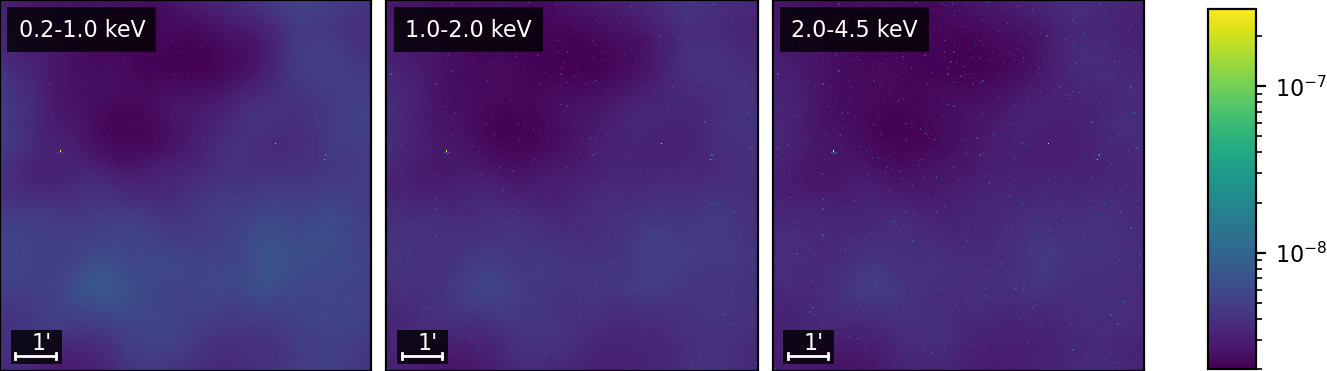} 
\caption{Visualization of the reconstruction per energy bin (left: 0.2-1.0 keV, center: 1.0-2.0 keV, right: 2.0-4.5 keV) in $[1/(\arcsec^2 \times \  \mathrm{s})]$.}
  \label{fig:mock_rec_ebin}
\end{figure*}

\begin{figure*}[!h]
\centering
  \includegraphics[width=0.94\linewidth]{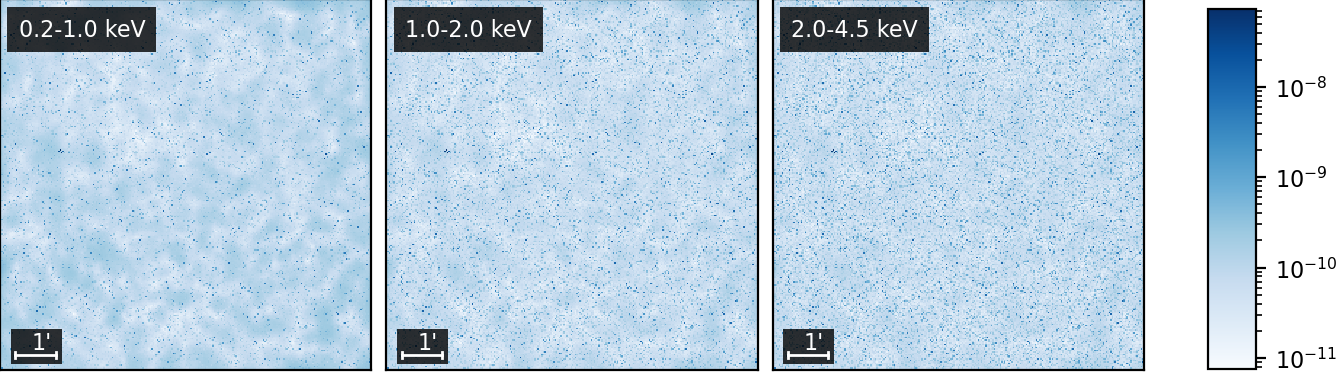}
\caption{Visualization of the standard deviation of the validation reconstruction per energy bin (left: 0.2-1.0 keV, center: 1.0-2.0 keV, right: 2.0-4.5 keV) in $[1/(\arcsec^2 \times \  \mathrm{s})]$.}
\label{fig:simulateduncertainty}
\end{figure*}

\begin{figure*}[!h]
\centering
  \includegraphics[width=0.94\linewidth]{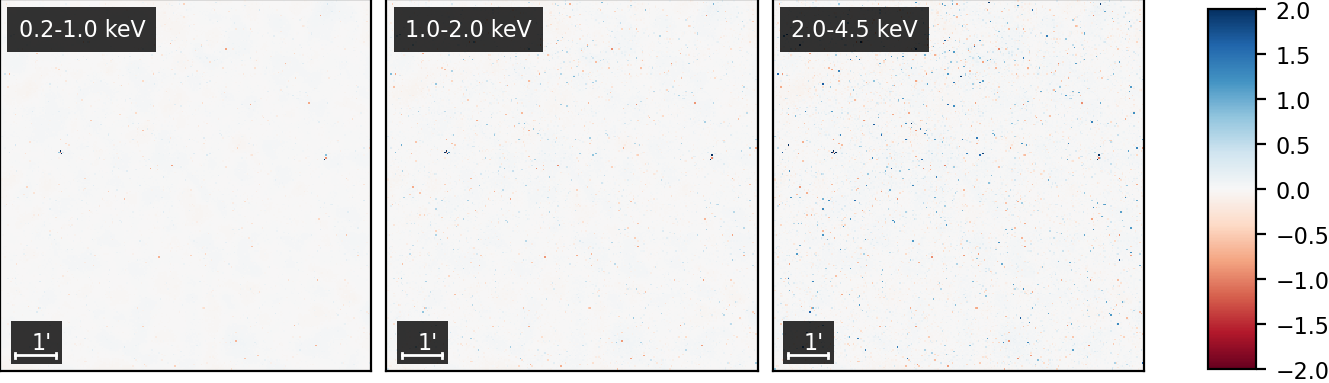}
\caption{Visualization of the standardized error of the validation reconstruction per energy bin (left: 0.2-1.0 keV, center: 1.0-2.0 keV, right: 2.0-4.5 keV).}
\label{fig:standardized error}
\end{figure*}

This section provides supplementary plots that offer further insights into the validation analysis discussed in Sect.~\ref{sec:validation}. These plots show in particular the images of the simulated sky, the simulated data, and the corresponding reconstruction.
These are shown as an RGB image in Fig.~\ref{fig:validation} as plots per energy, which serve to enhance the understanding of the color bar for the RGB image. In particular the data per energy bin and \ac{TM} is presented in Fig.~\ref{fig:simdata_ebin}, the underlying simulated sky per energy bin is shown in Fig.~\ref{fig:simsky_ebin} and the reconstructed sky per energy bin is illustrated in Fig.~\ref{fig:mock_rec_ebin}. Furthermore, we display the uncertainty in the reconstruction by means of the standard deviation for each energy bin in Fig.~\ref{fig:simulateduncertainty}.
Fig.~\ref{fig:standardized error} shows the standardized error for each energy bin.


\end{appendix}

\end{document}